\documentstyle[prb,aps,eqsecnum,multicol,psfig,array]{revtex}
\begin{document}
\title{Suppression of chaotic dynamics and localization of
two-dimensional electrons by a weak magnetic field}
\author{M. M. Fogler, A. Yu. Dobin\cite{StP}, V. I. Perel\cite{StP},
and B. I. Shklovskii}

\address{Theoretical Physics Institute, University of Minnesota,
116 Church St. Southeast, Minneapolis, Minnesota 55455}

\maketitle

\begin{abstract}

We study a two-dimensional motion of a charged particle in a weak random
potential and a perpendicular magnetic field. The correlation length of
the potential is assumed to be much larger than the de Broglie
wavelength. Under such conditions, the motion on not too large length
scales is described by classical equations of motion. We show that the
phase-space averaged diffusion coefficient is given by Drude-Lorentz
formula only at magnetic fields $B$ smaller than certain value $B_c$. At
larger fields, the chaotic motion is suppressed and the diffusion
coefficient becomes exponentially small. In addition, we calculate the
quantum-mechanical localization length as a function of $B$ in the
minima of $\sigma_{xx}$. At $B < B_c$ it is exponentially large but
decreases with increasing $B$. At $B > B_c$, the localization length
drops precipitously, and ceases to be exponentially large at a field
$B_\ast$, which is only slightly above $B_c$. Implications for the
crossover from the Shubnikov-de Haas oscillations to the quantum Hall
effect are discussed. 

\end{abstract}
\pacs{PACS numbers: 73.40.Hm, 80, 73.20.F}

\begin{multicols}{2}

\section{Introduction}
\label{intro}

In this paper we study a two-dimensional motion of a charged particle in
a weak random potential and a perpendicular magnetic field. This problem
has deep historical roots and the limiting cases of a weak and a very
strong magnetic field are fairly well understood. As we will see below,
the nature of the motion in these two limits is crucially different.
Surprisingly, until now no theory for the cross-over between the two
limits has been proposed. Our goal is to develop such a theory. We will
start with a classical description of the transport.

An important prediction of the classical magnetotransport theory is that
the conductivity in the direction perpendicular to the magnetic field is
reduced,
\begin{equation}
          \sigma_{xx} = \frac{\sigma_0}{1 + (\omega_c \tau)^2},
\label{sigma_Drude}
\end{equation}
where $\sigma_0$ is the zero field conductivity (the magnetic field $B$
is assumed to be along the $\hat{\bbox{z}}$-direction), $\omega_c = e B / m
c$ is the cyclotron frequency, and $\tau$ is the transport time
determined by the properties of the random potential. Strictly speaking,
in classical theory it is more consistent to study the diffusion
coefficient $D$. So, we would write Drude-Lorentz
formula~(\ref{sigma_Drude}) in the form
\begin{equation}
              D = \frac{D_0}{1 + (\omega_c \tau)^2},
\label{Drude}
\end{equation}
where $D_0 = \frac12 v^2 \tau$ is the diffusion coefficient in zero
field, $v$ being the particle velocity. Drude-Lorentz formula predicts
that if the magnetic field is not too weak so that $\omega_c \tau > 1$,
then the diffusion coefficient falls off inversely proportional to the
square of the magnetic field.

Let us examine the physical picture of the motion in such magnetic
fields. It is easy to verify that the Lorentz force has a
dominant effect on the motion and the deviations from the perfectly
circular cyclotron orbit are small. In such circumstances, the original
coordinates $\bbox{r} = (x, y)$ are not very useful any more. Instead,
it is convenient to study the motion of the guiding center $\bbox{\rho}
= (\rho_x, \rho_y)$ of the cyclotron orbit.

Suppose the cyclotron gyration is clockwise (this is the case if, e.g.,
the particle charge is negative and the magnetic field is in the
negative $\hat{\bbox{z}}$-direction). The guiding center coordinates
are defined as follows,
\begin{equation}
\rho_x = x + \frac{v_y}{\omega_c},\:\:\:
\rho_y = y - \frac{v_x}{\omega_c}.
\label{gc}
\end{equation}
Drude-Lorentz formula~(\ref{Drude}) results from the assumption that
the guiding center $\bbox{\rho}$ performs a random walk. The
characteristic step of such a random walk is the cyclotron radius, $R_c
= v / \omega_c$, and the time interval between the steps is the
transport time $\tau$. As we will see below, this is the
correct description of the motion if the magnetic field is not too strong.

Perhaps the first work that demonstrated that Drude-Lorentz formula may
not be valid in the limit of strong magnetic field was that of
Alfv\'en~\cite{Alfven} where he studied the motion of a
charged particle in an inhomogeneous electromagnetic field. This and
subsequent study~\cite{Drift_3d,Kruskal,Books} has led to the
recognition that instead of the random walk, the guiding center performs
a slow adiabatic drift along some well defined contours. The attention
to this problem was stimulated by its plasma physics applications, and
mostly three-dimensional case was considered. Not so long ago, the
extension to the two-dimensional case has been proposed by several
authors~\cite{Drift_2d} motivated by the quantum Hall effect
studies.~\cite{Prange} We will discuss the two-dimensional case from now
on.

Conventionally, the drift approximation is applied to the regime where the
magnetic fields so strong that the cyclotron radius $R_c = v / \omega_c$
is smaller than the correlation length $d$ of the random potential. In
this case the guiding center performs a drift along the constant energy
contours of the random potential. For the potential of a general type
all of such contours except one are closed loops and thus the motion is
finite. The motion is infinite only when a guiding center happens to be on the
so-called percolating contour.~\cite{Drift_2d} If one takes the drift
picture literally, and attempts to calculate the average diffusion
coefficient, the result will be equal to zero because the percolating
contour has zero measure.

Certainly, it has been understood that the drift picture is only an
approximation.
Nevertheless, the diffusion coefficient should be significantly smaller
than the Drude-Lorentz result~(\ref{Drude}). We will show that the
diffusion coefficient is, in fact, {\em exponentially small\/}.

Comparing the transport properties in the two regimes described above,
we see that the increase in the magnetic field drives the system from
the essentially delocalized, chaotic regime to the regime where the
motion is regular and the trajectories of the particles are localized.
We call this phenomenon the ``classical localization.'' The classical
localization occurs because of an extremely ineffective energy exchange
between two degrees of freedom of the particle, the cyclotron motion and
the guiding center motion. Without such an exchange the guiding center
is bound to a certain constant energy contour. At the same time, the
energy exchange is suppressed because the two degrees of freedom have
very different characteristic frequencies, the cyclotron frequency
$\omega_c$ being much larger than the drift frequency $\omega_d$.
Naturally, the present problem is directly related to the problem of a
nonconservation of adiabatic invariants. The latter is known to be
exponentially small,\cite{Adiabatic invariants} and therefore, it is
not so surprising that the diffusion coefficient turns out to be
exponentially small as well. 

One of the quantities we calculate in this paper is the value $B_c$ of
the magnetic field where the diffusion gives in to the classical
localization as $B$ increases. A naive guess would be the field where
$R_c = d$. Let us, however, compare $\omega_c$ and $\omega_d$ at such a
field. Denote the amplitude of the random potential $U(\bbox{r})$ by
$W$. We will assume that the potential is weak, $W \ll E$, where $E = m
v^2 / 2$ is the particle's energy. The characteristic drift velocity is
$v_d \sim \nabla U / m \omega_c \sim W / m\omega_c d$, and the drift
frequency $\omega_d \sim v_d / d \sim W / m \omega_c d^2$. Hence, the
ratio $\gamma$ of the two frequencies is
\begin{equation}
\gamma = \frac{\omega_d}{\omega_c} \sim \frac{W}{m \omega_c^2 d^2}
\sim \frac{W}{E} \left(\frac{R_c}{d}\right)^2,\quad R_c \lesssim d,
\label{gamma1}
\end{equation}
We see that at the point where $R_c = d$, this ratio is of the order of
$W / E \ll 1$. Surprisingly, the classical localization must first arise
already when $R_c \gg d$. To understand what kind of drift takes place
in this case one can use the averaging method. This method was
extensively developed by Krylov, Bogolyubov, and
Mitropolsky~\cite{Krylov} and in application to the problem at hand by
Kruskal.~\cite{Kruskal} In the spirit of this method, one has to imagine
that the slowly moving guiding center is entirely ``frozen'' on the time
scale of the cyclotron period. One then calculates the average potential
\begin{equation}
U_0(\rho_x, \rho_y) = \oint\frac{d \phi}{2 \pi} U(\rho_x + R_c \cos \phi,
    \rho_y + R_c \sin \phi),
\label{U_0} 
\end{equation}
acting on the particle during one cyclotron rotation. According to the
averaging method, the drift of the guiding center is performed along the
constant energy contours of the averaged potential $U_0(\bbox{\rho})$.
This conclusion was previously reached by Laikhtman.~\cite{Laikhtman} If
$R_c \ll d$, the average potential coincides with the bare one and so,
in agreement with the previous studies, the drift is performed along the
constant energy contours of the bare potential. However, if $R_c \gg d$,
then $U_0$ differs from $U$. The averaging reduces the amplitude of the
potential by a factor $\sqrt{R_c / d}$, which is the square root of the
number of uncorrelated ``cells'' of size $d$ along the cyclotron orbit
of length $2 \pi R_c$. Hence, $U_0$ has the amplitude $W_0 \sim W
\sqrt{d / R_c}$.

Now we can find the true boundary $B_c$ of the classical localization.
To this end we have to replace $W$ by $W_0$ in Eq.~(\ref{gamma1}), which
gives
\begin{equation}
\gamma \sim \frac{W}{E} \left(\frac{R_c}{d}\right)^{\frac32},
\quad R_c \gtrsim d,
\label{gamma2}
\end{equation}
and then solve $\gamma = 1$ for $B$. The result is
\begin{equation}
  B_c = \frac{\sqrt{m c^2 E}}{e d} \left(\frac{W}{E}\right)^{\frac23}.
\label{B_c}
\end{equation}
The change of the transport regime at such field was predicted earlier
by Baskin {\it et al.\/}~\cite{Baskin} and by
Laikhtman.~\cite{Laikhtman} These authors noted that the displacement
$\delta r$ of the guiding center after one cyclotron period is a
decreasing function of the magnetic field, $\delta r \sim \gamma d$ in
our notations. Thus, at $B > B_c$ where $\gamma < 1$, such a
displacement is smaller than the correlation length of the random
potential. As a result, the scattering by the potential is no longer a
sequence of uncorrelated acts and the motion of the guiding center is
different from the random walk, which invalidates Eq.~(\ref{Drude}).

Although the cross-over point $B_c$ has been identified correctly, the
understanding of the transport regime at larger magnetic fields remained
not entirely satisfactory. For example, Baskin {\it et
al.\/}~\cite{Baskin} arrived at a strange conclusion that at $B > B_c$
the diffusion coefficient becomes larger than that given by
Drude-Lorentz formula~(\ref{Drude}). On the other hand, the calculation
of Laikhtman~\cite{Laikhtman} relies on the existence of the random
inelastic scattering processes. In this paper we address the question of
{\it zero\/} temperature transport where all the scattering acts are due
to the static random potential only.

The key point of our approach is that the drift picture is albeit
excellent but an approximation. A more accurate analysis given in
Sec.~\ref{strong} reveals that the diffusion occurs not only when the
guiding center is situated precisely on the percolating contour but also
within a strip of finite width, so-called stochastic
layer,~\cite{Zaslavsky} surrounding this contour. Such a layer turns out
to be exponentially narrow if the magnetic field is larger than $B_c$.
As a result, the phase-space averaged diffusion coefficient $D$ is also
exponentially small,
\begin{equation}
                D \propto \omega_c d^2 e^{-B / B_c},
\label{D exp}
\end{equation}
Thus, the ``classical localization'' above $B_c$ causes strong
deviations from the conventional Drude-Lorentz formula~(\ref{Drude}).

The existence of the stochastic layer around the percolating contour is
quite natural. Indeed, the classical localization is owing to the fact
that drift trajectories are closed loops. It turns out that the drift
along the loops passing sufficiently close to the saddle points of the
random potential is unstable. The instability is realized as a slow
diffusion of the guiding center in the direction transverse to the drift
velocity. Suppose that the percolation level is $U_0 = 0$. By virtue of a
small transverse displacement, the particle drifting along the contour
$U_0 = -\epsilon$ can move to another closed contour $U_0 = +\epsilon$.
Although this displacement may be small, it will, in fact, lead to a
much larger displacement at a later time because the center of the other
loop is typically located a large distance away. Eventually, the
particle can travel infinitely far from its initial position. This
is the nature of the diffusion mechanism inside the stochastic layer.
This mechanism is guaranteed to exist because the percolating contour
necessarily passes through the saddle points.

The suppression of chaotic motion with increasing magnetic field
proceeds as follows. At $B < B_c$ the chaotic motion takes place in the
majority of the phase space, while the the regular motion is restricted
to small stability islands.~\cite{Zaslavsky} In this regime the
correlations among the scattering acts can be ignored and
Eq.~(\ref{sigma_Drude}) applies. As the magnetic field increases, the
regions of regular motion expand while the stochastic layer shrinks. Above
$B_c$ the width of the stochastic layer starts to decrease
exponentially leading to formula~(\ref{D exp}).

\begin{figure}
\centerline{
\psfig{file=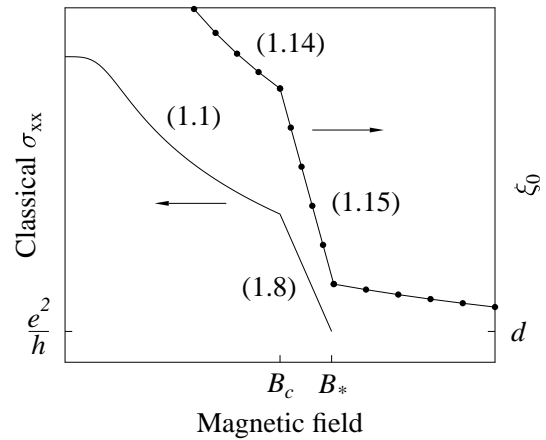,width=2.9in,bbllx=150pt,bblly=370pt,bburx=480pt,bbury=650pt}
}
\vspace{0.1in}
\setlength{\columnwidth}{3.2in}
\centerline{\caption{
Classical conductivity $\sigma_{xx}$ (solid line) and the localization
length $\xi_0$ at the {\it QHE conductivity minima\/} (solid line with
dots) as functions of the magnetic field (schematically). Dots serve as
a reminder that $\xi_0$ is defined at discreet values of the magnetic
field. The curves are labeled by the equation numbers, which render
their functional form in the corresponding intervals.
\label{Plot xi}
}}
\end{figure}

So far, we have discussed a purely classical dynamics. One can also
study the transport properties of a non-interacting electron system
quantum-mechanically. Due to quantum interference, the conductivity
of such a system turns out to be length-scale-dependent.~\cite{Lee_Rama}
The knowledge of classical dynamics enables one to find ``classical''
$\sigma_{xx}$, i.e., the conductivity, which would be
measured on not too large length scales where effects of quantum
interference are weak. Classical $\sigma_{xx}$ is calculated as a
product of the classical diffusion coefficient $D$ and the quantum
density of states $m / \pi \hbar^2$. It is given by Drude-Lorentz
formula~(\ref{sigma_Drude}) at $B < B_c$ and by a different formula
\begin{equation}
\sigma_{xx} \sim \frac{e^2}{h} G e^{-B / B_c},\quad
G = k_F d \left(\frac{W}{E}\right)^{\frac23},
\label{sigma exp}
\end{equation}
at $B > B_c$. Here $k_F = \frac{1}{\hbar}\sqrt{2 m E}$ is the Fermi
wavevector ($E$ has the meaning of the Fermi energy). The dependence of
classical $\sigma_{xx}$ on $B$ is illustrated graphically by
Fig.~\ref{Plot xi}. As one can see, classical $\sigma_{xx}$ quickly
drops above $B = B_c$. In Fig.~\ref{Plot xi} we indicated one special
value of the magnetic field, $B_\ast$, at which classical $\sigma_{xx}$
reaches $\frac{e^2}{h}$,
\begin{equation}
                             B_\ast = B_c \ln G.
\label{B_ast}
\end{equation}
Here we assume that $G \gg 1$, i.e., that
\begin{equation}
            d \gg k_F^{-1} \left(\frac{E}{W}\right)^{\frac23}.
\label{range}
\end{equation}

As we will see below this value of the magnetic field plays an important
role in the quantum transport.

%
%

At this point we would like to remind the reader that the true
$\sigma_{xx}$, i.e., the one which is measured experimentally, is the
conductivity on a large length scale (of the order of the sample size).
The calculation of this quantity is much more difficult. Similar to the
classical transport theory, there exist two mutually contradicting
approaches. One is the theory of the Shubnikov-de Haas (SdH) effect,
which aspires to predict the behavior of $\sigma_{xx}$ in weak magnetic
fields. The other is the theory of the quantum Hall effect (QHE), which
is conventionally applied to strong fields.

At present, the transition from the SdH regime to the QHE is not well
understood even for a non-interacting system. The
traditional explanation of the QHE is based on the idea of localization;
{\it viz.,\/} it is believed that at zero temperature an electron can
propagate diffusively only if its energy is precisely at the center of a
Landau level (in strong fields).~\cite{Prange} This leads to isolated
peaks in $\sigma_{xx}$, which are the
signature of the QHE. On the other hand, in the theory of the
SdH effect,~\cite{Ando,Laikhtman94} the suppression of $\sigma_{xx}$ is
related merely to the dips in the density of states between neighboring
Landau levels, while the idea of localization is totally discarded. This
crucial difference leads to different predictions for the conductivity
minima. Arguing from the QHE standpoint, one expects zero dissipative
conductivity, whereas the theory of SdH effect predicts a finite one.

In this paper we will advocate the following way to resolve this
apparent contradiction. We will argue that at the QHE conductivity
minima the states at the Fermi level are localized. At $B < B_\ast$ where
$B_\ast$ is given by Eq.~(\ref{B_ast}), the localization length $\xi_0$
of such states is {\it exponentially large\/} but decreases from one
minima to the next as $B$ increases. Above $B_c$ the fall-off of $\xi_0$
is extremely sharp and at $B \simeq B_\ast$, which is only
logarithmically larger than $B_c$, the localization length stops being
exponentially large. Consequently, $B = B_\ast$ is the smallest magnetic
field at which the observability of the QHE does not require {\em
exponentially small} temperatures. This fact motivate us to identify the
field $B = B_\ast$ as the starting point of the QHE. In other words,
this is the position of the ``first'' QHE plateau.

To avoid confusion let us further elaborate on this issue. Precisely at
zero temperature one will observe the QHE peaks. Between the peaks
$\sigma_{xx}$ will be exactly zero because of the quantum localization.
At finite temperature $T > 0$ inelastic processes appear, which break
the quantum coherence on length scales exceeding some
temperature-dependent length $L_\phi(T)$. Thus, if $\xi_0 > L_\phi(T)$,
then the quantum localization is not important and the QHE features
disappear. It is believed that the dependence of $L_\phi$ on $T$ is some
power law.~\cite{Huckestein} Therefore, if $\xi_0$ is exponentially
large, then the inequality $\xi_0 > L_\phi(T)$ is met already at
exponentially small temperatures.

There is yet another way to see why the observability of the QHE require
small $T$ when $\xi_0$ is large. It is known from experiment~(see the
bibliography of Ref.\onlinecite{Polyakov}) that the low-temperature
magnetotransport data at the $\sigma_{xx}$ minima is consistent
with the law
\begin{equation}
\sigma_{xx} \propto e^{-\sqrt{T_0 / T}},
\label{T_one_half}
\end{equation}
which can be interpreted~\cite{Polyakov} in terms of the variable-range
hopping in the presence of the Coulomb gap.~\cite{ES} In this theory
$T_0$ is directly related to $\xi_0$,
\begin{equation}
           T_0 = {\rm const}\,\frac{e^2}{\kappa \xi_0},
\label{T_0}
\end{equation}
where $\kappa$ is the dielectric constant of the medium. Deep minima of
$\sigma_{xx}$ are observable only if $T \ll T_0$. Thus, if $\xi_0$ is
exponentially large, then the QHE can be observed only at exponentially
small $T$. So, we reiterate once more that in practical terms there
exists a starting point of the QHE. The precipitous drop of $\xi_0(B)$
above $B_c$ leaves only a minimal ambiguity in identifying this point
with $B = B_\ast$.

Our calculation of the localization length $\xi_0$ at the QHE minima of
$\sigma_{xx}$ is based on the following {\it
ansatz\/},~\cite{Huckestein,Wei} which we discuss in more detail in
Sec.~\ref{quantum},
\begin{equation}
        \xi_0 \propto \exp(\pi^2 g_0^2), \quad g_0 \gg 1.
\label{xi_0}
\end{equation}
Here $g_0 = \frac{h}{e^2}\sigma_{xx}$ is the dimensionless classical
conductance. Using Eqs.~(\ref{sigma_Drude}) and (\ref{sigma exp}), we
immediately find
\begin{eqnarray}
\displaystyle \xi_0 &\propto& \exp\left(G^2 \frac{B_c^4}{B^4}\right),
\quad B_c \left({W}/{E}\right)^{\frac43} < B < B_c,
\label{xi_0 weak}\\
\displaystyle \xi_0 &\propto& \exp\left(G^2 e^{-2 B / B_c}\right),
\quad B_c < B < B_\ast,
\label{xi_0 strong}
\end{eqnarray}
The low-field end of the
interval in Eq.~(\ref{xi_0 weak}) corresponds to $\omega_c\tau \sim 1$.

As one can see from Eqs.~(\ref{xi_0 weak}) and (\ref{xi_0 strong}), the
localization length indeed drops precipitously above $B = B_c$. At $B =
B_\ast$, which is only logarithmically larger than $B_c$, $g_0$ becomes
of the order of unity and $\xi_0$ ceases to be exponentially large. The
dependence of $\xi_0$ on $B$ in the interval $B_c
\left({W}/{E}\right)^{\frac43} < B < B_\ast$ is illustrated by
Fig.~\ref{Plot xi}. The dependence of $\xi_0$ on $B$ at even stronger
magnetic fields, $B > B_\ast$, will be discussed in a forthcoming paper.
At this point we can only say that at such fields the localization
length is determined mainly by quantum tunneling and exhibits a power
law dependence on $B$.

In order to verify our predictions concerning $\xi_0(B)$ experimentally,
one has to measure $\sigma_{xx}$ at very low temperatures and fit the
data to the form~(\ref{T_one_half}). From such a fit one can deduce
$T_0$, which is directly related to $\xi_0$, see Eq.~(\ref{T_0}).

The paper is organized as follows. In Sec.~\ref{strong} we discuss the
classical dynamics in strong ($B \gg B_c$) magnetic fields and
demonstrate that the diffusion coefficient is exponentially small. In
Sec.~\ref{amplitudes} we analyze the same problem from the
quantum-mechanical point of view. Sec.~\ref{quantum} is devoted to the
calculation of the quantum localization length both in strong and weak
magnetic fields. Finally, in Sec.~\ref{conclusions} we summarize our
findings and discuss their relation to the experiment.

\section{Classical dynamics in strong magnetic fields}
\label{strong}

\begin{figure}
\centerline{
\psfig{file=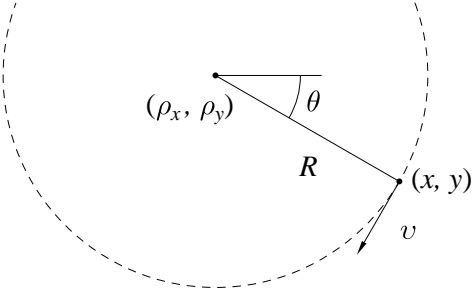,width=2.5in,bbllx=165pt,bblly=305pt,bburx=485pt,bbury=510pt}
}
\vspace{0.1in}
\setlength{\columnwidth}{3.2in}
\centerline{\caption{
The guiding center and cyclotron motion coordinates.
\label{def}
}}
\end{figure}

In this Section we study the classical dynamics of the system
with the Hamiltonian
\begin{equation}
H = \frac{(\bbox{p} + \frac{e}{c} \bbox{A})^2}{2 m} + U(\bbox{r}),
\quad \bbox{A} = (0, -Bx, 0).
\label{Hamiltonian}
\end{equation}
It corresponds to a particle with negative charge $-e$ and the magnetic
field in the negative $\hat{z}$-direction. Thus, the cyclotron gyration
is clockwise. By means of the canonical transformation with the
generating function
\[
F(x, y, \theta, \rho_y) = m \omega_c\left[x (y - \rho_y)
+ \frac{(y - \rho_y)^2}{2} \cot \theta\right]
\]
we obtain new momenta $-\partial F /\partial \rho_y = m \omega_c
\rho_x$ and $-\partial F/ \partial \theta \equiv I$. In terms
of the new variables, the Hamiltonian~(\ref{Hamiltonian}) acquires
the following form
\begin{equation}
H = I \omega_c + U(\rho_x + R \cos\theta, \rho_y - R \sin\theta),
\quad R \equiv \sqrt{\frac{2 I}{m \omega_c}}.
\label{Hamiltonian II}
\end{equation}
It is easy to see that the pair $(\rho_x, \rho_y)$ matches the earlier
definition~(\ref{gc}) of the guiding center coordinates. The
geometrical meaning of the other variables is illustrated by
Fig.~\ref{def}.

%
%

The equations of motion are
\begin{eqnarray}
&\displaystyle \dot\rho_x = -\frac{1}{m\omega_c}
\frac{\partial U}{\partial \rho_y},\:\:
\dot\rho_y = \frac{1}{m\omega_c}
\frac{\partial U}{\partial \rho_x},&
\label{rho_dot}\\
&\displaystyle \dot\theta = \omega_c + \frac{\partial U}{\partial I},\:\:
\dot{I} = -\frac{\partial U}{\partial\theta}.&
\label{theta_dot}
\end{eqnarray}
This system contains four dynamical variables, which makes its solution
difficult. We can eliminate one of the
variables, e.g., $I$, using the energy conservation. To this end we need
to solve the equation
\[
              E = I \omega_c + U(\bbox{\rho},\theta, I)
\]
for $I$, or equivalently, the equation
\[
           R^2 = \frac{2}{m\omega_c^2}[E - U(\bbox{\rho},\theta, R)]
\]
for $R$. For the potential $U$ of an arbitrary strength this can be
quite cumbersome. However, at least when the amplitude $W$ of
potential $U$ is small enough,
\begin{equation}
                            W \ll E \frac{R}{d},
\label{weakness}
\end{equation}
it is sufficient to use an approximate solution
\[
              R \simeq R_c \equiv \sqrt{\frac{2 E}{m \omega_c^2}}.
\]
Condition~(\ref{weakness}) guarantees that the deviation of $R$
from $R_c$ is much smaller than the correlation length $d$ of
potential $U$. Under this condition we can also neglect the deviation of
$\dot\theta$ from $\omega_c$. As a result,
Eqs.~(\ref{rho_dot}) and (\ref{theta_dot}) can be treated as the equations of
motion for the time-dependent Hamiltonian
\begin{equation}
        H = U(\rho_x + R_c\cos\omega_c t, \rho_x - R_c\sin\omega_c t)
\label{H new}
\end{equation}
with $\rho_y$ being the canonical coordinate and $m\omega_c\rho_x$ being
the canonical momentum. It is customary to classify the systems of this
kind as systems with $1\frac12$ degrees of freedom. 

It is useful to expand Hamiltonian~(\ref{H new}) in the Fourier series,
\begin{equation}
            H = \sum_k U_k(\bbox{\rho}) e^{-i k \omega_c t},
\label{Hamiltonian III}
\end{equation}
with the expansion coefficients given by
\begin{equation}
U_k(\bbox{\rho}) \equiv \oint \! \frac{d \phi}{2 \pi}
  U(\rho_x + R_c \cos \phi, \rho_y + R_c \sin \phi) e^{-i k \phi},
\label{U_k} 
\end{equation}
[compare with Eq.~(\ref{U_0})]. The new equation of motion for
$\rho_x$ is
\begin{equation}
\dot\rho_x = -\frac{1}{m\omega_c}\frac{\partial U_0}{\partial \rho_y}
 - \frac{1}{m\omega_c} \sum_{k\neq 0}
 \frac{\partial U_k}{\partial \rho_y}\, e^{-i k \omega_c t},
\label{rho_x_dot}
\end{equation}
and similarly for $\dot\rho_y$. If we drop the sum on the right-hand
side of Eq.~(\ref{rho_x_dot}), then the remaining term will describe the
drift of the guiding center along the contours of constant $U_0$. Such a
drift leads to the classical localization described in the previous
Section. The characteristic drift frequency is of the order of $\omega_d
\sim W_0 / m \omega_c^2 d^2$, where $W_0$ is the amplitude of $U_0$ (see
Sec.~\ref{intro}). Thus, if the parameter $\gamma = \omega_d / \omega_c$
is small, then all the terms in the sum on the right-hand side
Eq.~(\ref{rho_x_dot}) have frequencies much larger than the $\omega_d$.
They can be considered a high-frequency perturbation imposed on the
``unperturbed'' drift motion.

The presence of a small parameter calls for the perturbation theory
treatment (averaging method) developed in
Refs.~\onlinecite{Drift_3d,Kruskal,Books}. Unfortunately, it is not
possible to calculate the diffusion coefficient perturbatively because
the perturbation theory series converge only asymptotically, i.e., they
formally diverge for any finite $\gamma$. The calculation of the
diffusion coefficient requires a different approach based on the
consideration of the chaotic dynamics of the system within a narrow
stochastic web surrounding the percolating contour of the potential
$U_0(\bbox{\rho})$.

Due to an extreme difficulty of the problem, we restrict our
consideration by two particular examples: a chessboard potential
and a Gaussian random potential.

\subsection{Chessboard geometry}
\label{chess}

Consider a chessboard potential
%
\[
U(x, y) = -W \left(\cos\frac{x}{d} + \cos\frac{y}{d}\right).
\]
%
In this case $U_0$ is given by
\begin{equation}
       U_0 = -W {\cal J}_0\left(R_c / d\right)
         \left(\cos\frac{\rho_x}{d} + \cos\frac{\rho_y}{d}\right).
\label{U_0 cb}
\end{equation}
More generally,
%
\[
       U_k = -W {\cal J}_k\left(\frac{R_c}{d}\right) \left\{
\begin{array}{cr}
\displaystyle i^k\cos\frac{\rho_x}{d}+\cos\frac{\rho_y}{d},&{\rm even}\,k\\[12pt]
\displaystyle i(i^k\sin\frac{\rho_x}{d}+\sin\frac{\rho_y}{d}),&{\rm odd}\,k,
\end{array}\right.
\]
%
where ${\cal J}_k$'s are the Bessel functions. 

As explained above, one can introduce the dimensionless parameter
$\gamma$, which governs the classical dynamics. Equation~(\ref{U_0 cb})
suggests that the appropriate definition for $\gamma$ is
%
\[
       \gamma = \frac{W}{m \omega_c^2 d^2} |{\cal J}_0(R_c / d)|.
\]
%

Note that with this definition $\gamma$ vanishes whenever $R_c / d$
coincides with a zero of ${\cal J}_0$. This property is a peculiarity of
the periodic geometry and it leads to oscillations in the diffusion
coefficient with the magnetic field, which are well known to exist both
from theory and from experiment.~\cite{Weiss,Fleishmann} This behavior
is nonuniversal and therefore, is not of primary interest to us. In the
following we will assume that the ratio $R_c/d$ is always close to
midpoints between the successive zeros of ${\cal J}_0$. In this case,
the dependence of $\gamma$ on $R_c$ is given by
Eqs.~(\ref{gamma1}) and (\ref{gamma2}). We will focus on the case $\gamma \ll
1$.

\begin{figure}
\centerline{
\psfig{file=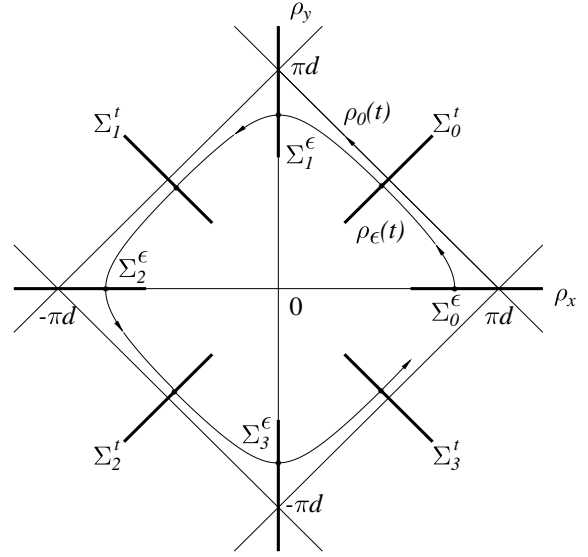,width=3.0in,bbllx=150pt,bblly=195pt,bburx=520pt,bbury=560pt}
}
\vspace{0.1in}
\setlength{\columnwidth}{3.2in}
\centerline{\caption{
A sketch illustrating the construction of the separatrix map. Two
unperturbed orbits, $\bbox{\rho}_0(t)$ and $\bbox{\rho}_\epsilon(t)$ are
shown. They follow two constant energy contours, $U_0 = 0$ (the
separatrix) and $U_0 = \epsilon < 0$, respectively. The energy-time
coordinates $\epsilon_n$ and $t_n$ are defined by the crossings of the
trajectories with the surfaces of section $\Sigma^\epsilon_q$ and
$\Sigma^t_q$ (shown by bold segments).
\label{sections}
}}
\end{figure}

The ``unperturbed'' motion is described by the Hamiltonian
%
\[
                     H_0 = U_0(\bbox{\rho}),
\]
%
which is time-independent. Hence, $U_0$ is the integral of motion in
agreement with the statement that the drift is performed along the
contours $U_0 = {\rm const}$. The motion has a periodic array of
hyperbolic (or saddle) points. Some of them, $(\pi d, 0)$, $(0, \pi d)$,
$(-\pi d, 0)$, $(0, -\pi d)$ are shown in Fig.~\ref{sections}, the
others can be obtained by periodic translations. The hyperbolic points
are connected by heteroclinic orbits or separatrices. One of them, which
runs from $(\pi d, 0)$ to $(0, \pi d)$ is shown in Fig.~\ref{sections}.
It has the following time dependence,
\begin{equation}
\begin{array}{l}
\displaystyle\rho_y = 2 d\arctan e^{\gamma\omega_c(t - t_0)},\\
\rho_x = \pi d - \rho_y,
\end{array}
\label{unperturbed}
\end{equation}
where $t_0$ is the moment of crossing the surface of section
$\Sigma^t_0$ (see Fig.~\ref{sections}). The heteroclinic orbits passing
through the other ``time surfaces'' $\Sigma^t_q$ (see
Fig.~\ref{sections}) have a similar functional form and an analogous
dependence on the crossing times $t_n$'s.

%
%

As explained in the Introduction, the unperturbed separatrix is dressed
with a narrow stochastic layer. In the case of the chessboard
potential, this layer has a topology of a square network. We are
interested in the long-time asymptotic behavior of the chaotic transport
along this network. An efficient tool to study such a transport is the
separatrix map.~\cite{Zaslavsky68,Chirikov} The separatrix map is an
approximate map describing the dynamics near the separatrix. The
application of the separatrix map to transport problems has been
previously considered in
Refs.~\onlinecite{Escande,Weiss_Knobloch,Lichtenberg_Wood,Afanasiev,Ahn}.

To construct the separatrix map we will consider ``energy surfaces''
$\Sigma^\epsilon_q$ in addition to the introduced above time surfaces
$\Sigma^t_q$. To avoid confusion we will elaborate a bit on the
definition of such surfaces. $\Sigma^\epsilon_q$'s and $\Sigma^t_q$'s are
introduced for each chessboard cell. Index $q$ runs from $0$ to $3$.
The energy surfaces come through the saddle points and the time surfaces
are drawn through the links connecting the neighboring saddle points.
The locations of $\Sigma^\epsilon_q$'s and $\Sigma^t_q$'s near the
perimeter of the cell at the origin are clear from Fig.~\ref{sections}.
The locations of the surfaces of section in the other cells can be
obtained by periodic translations. Thus, index $q$ in
$\Sigma^t_q$ refers to the position of the corresponding link with
respect to a given cell's center. Similarly, index $q$ in
$\Sigma^\epsilon_q$ refers to the position of the saddle point.
 
Let $\bbox{\rho}(t)$ be the exact trajectory near the separatrix. As $t$
increases, $\bbox{\rho}(t)$ crosses the surfaces $\Sigma^\epsilon_q$ in
certain order. We denote by $q_n$ the index of $\Sigma^\epsilon_q$ at
$n$-th crossing and by $\epsilon_n$ the value of $U_0$ at this moment.
Due to the time-dependent terms in the Hamiltonian, $\epsilon_n$
changes with $n$. Let us find the difference $\epsilon_{n + 1} - 
\epsilon_n$. The time derivative of $U_0$ is given by
\[
\frac{d U_0}{d t} = \frac{1}{m \omega_c}\sum_{k \neq 0}\left(
\frac{\partial U_0}{\partial \rho_y}\frac{\partial U_k}{\partial \rho_x}
-\frac{\partial U_0}{\partial \rho_x}\frac{\partial U_k}{\partial \rho_y}
\right) e^{-i k \omega_c t}.
\]
All $U_k$'s in this equation have to be calculated on the exact
trajectory $\bbox{\rho}(t)$, which is not known. Therefore,
following Refs.~\onlinecite{Zaslavsky,Zaslavsky68,Chirikov}, we perform
the following approximations. First we replace the exact trajectory by
the unperturbed one with $U_0 = \epsilon_n$. Second, having in mind that
$|\epsilon_n| \ll W_0$, we replace the trajectory with $U_0 =
\epsilon_n$ by the separatrix motion $\rho_0(t - t_n)$ where $\rho_0(t)$
is given by equations similar to Eq.~(\ref{unperturbed}) and $t_n$ is the
moment of time when $\bbox{\rho}(t)$ crosses the surface of section
$\Sigma^t_{q_n}$. As a result, we find
\begin{equation}
           \epsilon_{n + 1} - \epsilon_n = M_n(t_n),
\label{eeM}
\end{equation}
where $M_n$ is given by
\[
M_n(t) = \frac{1}{m \omega_c}\sum_{k \neq 0}
\int\limits_{-\infty}^\infty \!\!d t' \!\left(
\frac{\partial U_0}{\partial \rho_y}\frac{\partial U_k}{\partial \rho_x}
-\frac{\partial U_0}{\partial \rho_x}\frac{\partial U_k}{\partial \rho_y}
\right) e^{-i k \omega_c t'}
\]
and is termed the Melnikov function.~\cite{Melnikov} The integration is
done along the trajectory $\bbox{\rho}_0(t' - t)$. It can be shown that
the terms $k = \pm 1$ in the sum yield the dominant contribution. Thus,
the Melnikov function can be approximated by the integral
\begin{eqnarray}
\displaystyle M_n(t) &\simeq& 2 \gamma\omega_c W\: {\rm Re}\!\!
\int\limits_{-\infty}^\infty \! d t'\,
\frac{\tanh[\gamma\omega_c(t' - t)]}{\cosh[\gamma\omega_c(t' - t)]}
\nonumber\\
\displaystyle \mbox{} &\times& {\cal J}_1(R_c / d)\,\exp\!
\left(-i \omega_c t' + \frac{\pi}{4} + \frac{\pi q_n}{2}\right).
\label{M}
\end{eqnarray}
The integral can be evaluated by shifting the integration path to the
complex plane of $t$. Then $M_n(t)$ can be represented by the sum of
residues at the poles of the integrand. The residues from the poles
closest to the real axis dominate the sum. Retaining only these terms,
we arrive at
\begin{eqnarray}
&M_n(t) \simeq \Delta\epsilon \sin\vartheta_n,&
\label{M_n}\\
&\displaystyle \Delta\epsilon = 4\sqrt{2}\,\pi m \omega_c^2 d^2
\frac{{\cal J}_1(R_c / d)}{{\cal J}_0(R_c / d)}
e^{-\pi / 2 \gamma},&
\label{Delta E}\\
&\displaystyle \vartheta_n = \omega_c t_n + \frac{\pi}{4} +
\frac{\pi q_n}{2}.&
\label{vartheta}
\end{eqnarray}
Combining formulas~(\ref{eeM}) and (\ref{M_n})
we obtain the first equation of the separatrix mapping
\begin{equation}
\epsilon_{n + 1} = \epsilon_n + \Delta\epsilon \sin\vartheta_n.
\label{eps map}
\end{equation}
To have the mapping in a closed form we need another equation relating
$t_{n + 1}$ to $t_n$ and $\epsilon_n$. Following
Refs.\onlinecite{Zaslavsky,Zaslavsky68,Chirikov}, we take
\begin{equation}
           t_{n + 1} = t_n + \frac14 T(\epsilon_{n + 1}),
\label{t map}
\end{equation}
where $T(\epsilon)$ is the period of the unperturbed orbit
$U_0(\bbox{\rho}) = \epsilon$. A straightforward computation gives
\begin{eqnarray}
\displaystyle \frac{T(\epsilon)}{4} &=& \frac{1}{\gamma\omega_c}
K\left(1 - \frac{\epsilon^2}{4 W_0^2}\right)\nonumber\\
\displaystyle &\simeq& \frac{1}{\gamma\omega_c}
\ln \left|\frac{8 W_0}{\epsilon}\right|, \quad |\epsilon| \ll W_0,
\label{T_E}
\end{eqnarray}
$K$ being the complete elliptic integral of the first kind.

Although it is a common
practice~\cite{Zaslavsky,Zaslavsky68,Chirikov,Escande,Weiss_Knobloch,Lichtenberg_Wood,Afanasiev,Ahn,Melnikov}
to make the approximations similar to those we made above, their
validity is far from being obvious. The justification has come only
recently with a new development by Treschev.~\cite{Treschev} The
extension of Treschev's analysis to our problem~\cite{Comment on
Treschev} indicates that the naive calculation of the Melnikov function
is correct for $R_c \ll d$. If $R_c \gg d$, then Eq.~(\ref{Delta E}) is
off by a numerical factor and the replacement
\begin{equation}
\frac{{\cal J}_1(R_c / d)}{{\cal J}_0(R_c / d)} \rightarrow
j(R_c / d)
\label{j}
\end{equation}
is needed, where $j(x)$ is some function of the order of unity for all
real $x$.

Besides the analytical methods, the validity of the separatrix map
has been investigated numerically by several authors~\cite{Chirikov,Ahn}
and has been rated from ``satisfactory'' to ``excellent.'' In the rest of
this subsection we will assume that this is the case and calculate two
quantities relevant for the transport, the width $\Delta\epsilon_{\rm web}$
of the stochastic layer around the separatrix and the average
diffusion coefficient $D$.

We estimate $\Delta\epsilon_{\rm web}$ following
Ref.\onlinecite{Chirikov}. First, we note that the relative change in
$\epsilon_n$ after one application of the separatrix map is small
provided $|\epsilon_n| \gg \Delta \epsilon$. Under this condition the
map can be linearized. The defining parameter of the linearized map is
${\cal K}$, 
\[
{\cal K} \equiv \frac{1}{\cos\vartheta_n}\left(
    \frac{\partial\vartheta_{n + 1}}{\partial\vartheta_n} - 1\right)
  = \frac{\omega_c\Delta \epsilon}{4} \frac{d T(\epsilon_n)}{d \epsilon_n}
 = -\frac{\Delta \epsilon}{\gamma \epsilon_n},
\]
and the map itself coincides with the standard map.~\cite{Chirikov} The
crossover to the global stochasticity in the standard map occurs at
$|{\cal K}| \simeq 0.97$ [Ref.~\onlinecite{Greene}], which yields the
estimate
\begin{equation}
 \epsilon_{\rm web} \simeq 18 j(R_c / d) \frac{m \omega_c^2 d^2}{\gamma}
                        e^{-\pi /2 \gamma}
\label{width}
\end{equation}
for the stochastic layer's width. Note that $\epsilon_{\rm web} \sim \Delta
\epsilon / \gamma$ is much larger than $\Delta \epsilon$, and so the
approximation by the standard map is justified.

Let us now turn to the evaluation of the diffusion coefficient $D$. For
the chessboard geometry this problem has been considered previously by
Ahn and Kim.~\cite{Ahn} Unfortunately, they calculated the diffusion
coefficient averaged only over the trajectories inside the stochastic
layer. We, however, are interested in the diffusion coefficient averaged
over the {\em entire} phase space. Our approach to calculating $D$ is
close in spirit to the ones used for calculation of the diffusion
coefficient in planar periodic vortical flows, e.g., Rayleigh-B\'enard
cells.~\cite{Rosenbluth,Shraiman} The details of the calculation
can be found in Appendix~\ref{f_c}. The result is
\begin{equation}
             D = 0.45 \frac{\Delta \epsilon}{m \omega_c}.
\label{D I}
\end{equation}
Substituting this value into Eq.~(\ref{D I}), we obtain
\begin{equation}
    D = 7.9\,j\left({R_c}/{d}\right) \omega_c d^2\,e^{-\pi/2\gamma}.
\label{D_cb}
\end{equation}
Note that apart from the numerical factor, $D$ can be obtained from
following simple arguments. Consider an ensemble of particles moving in
the chessboard potential. Their diffusive motion can be visualized as a
random walk from one chessboard cell to the next. The motion of each
particle is a combination of the drift along the cell perimeter and the
series of random displacements in the transverse direction. The rate of
diffusion depends on the distance of a particle from the cell
boundaries. The particles located within a distance of one transverse
step from the cell boundaries possess the fastest rate because they can
cross to the neighboring cell after a single passage along the cell's
side. Particles further away from the perimeter remain trapped within
the same cell for much longer time. Hence, their diffusion rate is
negligible. Naturally, we can consider a model with an
$\epsilon$-dependent diffusion coefficient $D(\epsilon) =
\Theta(\Delta\epsilon - |\epsilon|)\, d_0^2 / T(\epsilon)$ where
$\Theta(x)$ is the step-function and $d_0 = \sqrt{2}\,\pi d$ is the
length of the cell's side. The net diffusion coefficient can be obtained
by averaging $D(\epsilon)$ over the phase space, i.e., over the area in
coordinates $(\rho_x, \rho_y)$,
\[
D = \frac{1}{d_0^2}\int\limits_0^{\Delta\epsilon} \!d\epsilon\,D(\epsilon)
\frac{d S(\epsilon)}{d\epsilon},
\]
where $S(\epsilon)$ is the area of the cell's region bounded by the
contours $U_0 = 0$ and $U_0 = \epsilon$. It is trivial to show that
$d S(\epsilon) / d\epsilon = T(\epsilon) / m\omega_c$; therefore, $D =
\Delta\epsilon / m\omega_c$, which reproduces Eq.~(\ref{D I}) up to a
numerical factor.

Finally, the diffusion coefficient can be written as a function of the
magnetic field $B$,
\begin{equation}
\ln \left(\frac{D}{\omega_c d^2}\right) \sim
-\left(\frac{B}{B_{\rm cb}}\right)^{3/2},
\label{D_of_B}
\end{equation}
where
\[
B_{\rm cb} = \frac{2^{7/6}}{\pi}\frac{\sqrt{m c^2 E}}{e d}
\left(\frac{W}{E}\right)^{2/3}
\]
[cf. Eq.~(\ref{B_c})]. Formula~(\ref{D_of_B}) was derived assuming that
$\gamma \ll 1$, i.e., that $B \gg B_{\rm cb}$. In addition, we assumed
that $R_c \gg d$, which is equivalent to $B \ll B_{\rm cb} \left({E} /
{W}\right)^{2/3}$. As one can see, the dependence of $D$ on $B$ for the
chessboard geometry is given by a squeezed exponential with the
exponent $\frac32$. In the next subsection we treat a more general case
of a Gaussian random potential. We will show that the squeezed
exponential is replaced by a simple one as given by Eq.~(\ref{D exp}).

\subsection{Gaussian random potential}
\label{random}

A Gaussian random potential is fully specified by its two-point
correlator $C(\bbox{r}_1 - \bbox{r}_2)$,
\begin{eqnarray*}
C(\bbox{r}_1 - \bbox{r}_2) &=&
\langle U(\bbox{r}_1) U(\bbox{r}_2) \rangle,\\
C(0) &\equiv& W^2.
\end{eqnarray*}
In many cases, it is also convenient to deal with the Fourier
transforms of $U$, which have the following correlator
\[
\langle \widetilde{U}(\bbox{q}_1) \widetilde{U}(\bbox{q}_2) \rangle =
(2 \pi)^2 \delta(\bbox{q}_1 + \bbox{q}_2)
\widetilde{C}(\bbox{q}_1)
\]
(Fourier transforms are denoted by tildes). Given the function
$C(\bbox{r})$, we want to calculate the diffusion coefficient in strong
magnetic fields. Similar to the case of the chessboard potential, let
us first investigate the ``unperturbed'' motion, the drift along the
contours $U_0(\bbox{\rho}) = {\rm const}$. Clearly, $U_0(\bbox{\rho})$
is also a Gaussian random potential with correlator $C_0$ related
to $C$ by
\[
 \widetilde{C}_0(q) = \left[{\cal J}_0(q R_c)\right]^2 \widetilde{C}(q).
\]

The unperturbed motion is determined by the properties of
the level lines of $U_0$. It is known that all of such lines except one,
the percolating contour, are closed loops. The Gaussian random potential
shares this property with the chessboard potential considered above. In
addition, the position of the percolation level is the
same for both potentials: $U_0 = 0$. There exists, however, an important
difference in the properties of level lines in the two cases. The
diameters of the loops in the chessboard do not exceed $2\pi d$. On
the other hand, constant energy contours of the random potential
can have arbitrarily large diameters. Such large
loops are found in the vicinity of the percolating contour. (The latter
one can be considered as a loop with infinitely large diameter). As the
diameter of the contour increases, the range of $U_0$ found at such
contours shrinks, tending to the percolation level $U_0 = 0$.

Similar to the chessboard geometry case, the exact trajectories do not
simply follow the level lines of $U_0(\bbox{\rho})$ but exhibit small
transverse deviations from them. As a result, a finite diffusion
coefficient appears. To calculate $D$ we will use a close analogy of the
problem at hand with the problem of calculating the effective diffusion
constant of a particle diffusing in an incompressible
flow.~\cite{Isichenko} Below we essentially reproduce the basic
arguments of Isichenko {\it et al.\/}\cite{Isichenko} with slight
modifications appropriate for our problem.

Borrowing the terminology of Ref.\onlinecite{Isichenko}, we call a
bundle of constant $U_0$ contours with diameters between $a$ and $2 a$ a
convection cell or an $a$-cell. The values of $U_0$ in typical $a$-cells
belong to an interval $[-w(a),\: w(a)]$, which narrows with increasing
$a$. Let us denote by $L(a)$ the perimeter length of typical $a$-cells
and by $\Delta\epsilon(a)$ the change in $U_0$ accumulated along the
trajectory following the perimeter, for which the time $T(a) \sim L(a) /
v_d$ is required. The key point in estimating $D$ is a ramification
between mixing [with $\Delta\epsilon(a) > w(a)$] and non-mixing
[$\Delta\epsilon(a) < w(a)$] cells. It takes a single period $T(a)$ or
even a fraction of thereof for the particle to leave a mixing cell,
whereas particles in non-mixing cells remain trapped for time intervals
much larger than $T(a)$. The dominant contribution to the transport
comes from the mixing cells of the largest width $w(a)$ for which
$\Delta\epsilon(a) \sim w(a)$. Denote the diameter by such cells by
$a_m$. The particles situated in such cells perform a random walk from
one optimal cell to the next. The characteristic step of the random walk
is $a_m$ and the characteristic rate of the steps is $1 / T(a_m)$. Thus,
the diffusion coefficient of such ``active'' particles is of the order
of $a_m^2 / T(a_m)$. The net diffusion coefficient can be found by
multiplying this diffusion coefficient by the fraction of the total area
occupied by the optimal convection cells. Note that the width of the
$a_m$-cells in the real space is of the order of $\Delta \epsilon(a_m) d
/ W_0$. Using this, the fraction of the area can be estimated to be
$[\Delta\epsilon(a_m) d / W_0] L(a_m) / a_m^2 = [\Delta\epsilon(a_m) / m
\omega_c] T(a_m) / a_m^2$. Finally, we obtain
\begin{equation}
           D \sim \frac{\Delta\epsilon(a_m)}{m\omega_c},
\label{D_de}
\end{equation}
which closely resembles Eq.~(\ref{D I}) for the
chessboard.~\cite{Comment on Isichenko} However, now $\Delta\epsilon_m
\equiv \Delta\epsilon(a_m)$ depends on the diameter $a_m$ of the optimal
cells, which has yet to be found. We see that the calculation of $D$
hinges upon the calculation of $\Delta\epsilon_m$. To accomplish the
latter task we can make the same kind of approximations as in deriving
the separatrix mapping for the chessboard. Then we obtain the following
expression,
\begin{eqnarray}
\Delta\epsilon_m^2 &=& \sum_{n \neq 0} |\Delta_n|^2,
\label{Delta epsilon}\\
\displaystyle \Delta_n &\equiv& \oint\limits \!d t\,
\bbox{v}_d \bbox{\nabla} U_n[\bbox{\rho}_0(t)]\,e^{-i n \omega_c t},
\label{Delta_n}
\end{eqnarray}
where the integration path is the unperturbed orbit
$U_0[\bbox{\rho}_0(t)] = {\rm const}$ belonging to a given $a_m$-cell.
Observe that the integrand is the product of a slowly changing function
$f_n(t) = \bbox{v}_d \bbox{\nabla} U_n[\bbox{\rho}_0(t)]$ and a rapidly
oscillating exponential factor $e^{-i n \omega_c t}$. It is customary to
estimate such integrals by shifting the integration path into the lower
half-plane of complex $t$ where the oscillating factor decays
exponentially. By using the method, one arrives at the following estimate
\begin{equation}
\Delta\epsilon_m^2 \sim |\Delta_1|^2 = \left|\sum_k 2\pi i R_k\,
e^{-|{\rm Im}\, \tau_k| \omega_c} \right|^2,
\label{Delta eps}
\end{equation}
where $\tau_k$ are the singular points of the function $f_1(t)$ in the
lower half-plane plane and $R_k$ are some pre-exponential factors. For
example, if $f_1(t)$ has a simple pole at $\tau_k$, then $R_k$ is up to a
phase factor the residue of such a pole. Equation~(\ref{Delta eps}) is
similar to Eqs.~(\ref{M_n}-\ref{vartheta}) for the chessboard
potential.

Denote the coordinate along the drift trajectory by $s$, then $f_1(t) =
v_d\,(d U_1 / d s)$. The singularities of $f_1(t)$ may originate either
from $v_d$ or from $(d U_1 / d s)$. Let us investigate the former
possibility. To get the necessary insight we will use the exactly
solvable model of the chessboard potential, which we studied above.
In the latter case
\begin{equation}
v_d(t) = \frac{\sqrt{2}\,\gamma\omega_c d}{\cosh[\gamma\omega_c(t - t_0)]}
\label{v_d}
\end{equation}
[see Eq.~(\ref{unperturbed})] and the singularities of $v_d(t)$ in the
lower half-plane consist of the ``parent'' pole at $t_0 - i\pi /
2\gamma\omega_c$ and a series of ``daughter'' poles at $t_0 - i \pi
\left(k + \frac12\right)/ \gamma\omega_c,\:\, k = 1,2\ldots$. Note that
the imaginary part of the parent pole is of the order of the
characteristic time scale $(\gamma\omega_c)^{-1}$ of the drift motion.

In the case of the random potential, we also expect to find series of
singularities of $v_d(t)$. However, there will be not a single series
but a large number $N(a_m)$ of them. Indeed, $v_d(t)$ has about $L(a_m)
/ d$ minima on the trajectory $s(t)$. The points of minima divide the
trajectory into $L(a_m) / d$ intervals of length $\sim d$. In each
interval $v_d(t)$ first rises, then reaches a maximum, then decreases,
i.e., it exhibits the same kind of behavior as in the chessboard case.
Therefore, a naive estimate of $N(a_m)$ is $N(a_m) \sim L(a_m) / d$.
Since ${\rm Im}\,\tau_k$'s enter Eq.~(\ref{Delta eps}) in the arguments
of the exponentials, the dominant contribution to $\Delta\epsilon_m$
comes from these $N(a_m)$ parent singularities. Let us now discuss ${\rm
Im}\, \tau_k$'s. It is obvious that different $a_m$-cells give rise to
different ${\rm Im}\, \tau_k$'s, i.e., there exists a certain
distribution of ${\rm Im}\, \tau_k$'s. What kind of distribution should
we expect? Clearly, the {\em typical} value of the imaginary parts of
the parent singular points should be of the order of the characteristic
time scale of the drift motion, $(\gamma\omega_c)^{-1}$, where $\gamma$
can be defined as follows:
%
\[
                \gamma = \frac{W_0}{m \omega_c^2 d^2},
\]
%
with $W_0$ and $d$ being
\[
   W_0 = \sqrt{C_0(0)},\:\: d = \sqrt{-\frac{C_0}{2 \bbox{\nabla}^2 C_0}}.
\]
However, it
would be a mistake to think that $\Delta\epsilon_m$ is determined by
this typical value. Indeed, the deviations of ${\rm Im}\, \tau_k$
from their average value are dramatically enhanced in
$\Delta\epsilon_m$. Therefore, we can expect an extremely broad range of
the exponential factors entering the sum on the right-hand side of
Eq.~(\ref{Delta eps}). At the same time, there is no such an enhancement
for $R_k$. This kind of arguments imply that we can estimate
$\Delta\epsilon_m$ considering only the distribution of ${\rm Im}\,
\tau_k$'s, i.e.,
\[
\Delta\epsilon_m^2 \sim \left|\sum_k e^{i\vartheta_k}
                  e^{-|{\rm Im}\, \tau_k| \omega_c}\right|^2,
\]
where $\vartheta_k$ is the phase of the complex number $R_k$. Of course,
the pre-exponential factor in $\Delta\epsilon_m$ can not be found by
this approach. We will further assume that $\vartheta_k$'s are
uncorrelated, which results in
\[
\Delta\epsilon_m^2 \sim \sum_k e^{-2 |{\rm Im}\, \tau_k| \omega_c}.
\]
From this, we find that
\begin{equation}
  \Delta\epsilon_m^2 \sim W_0^2 \frac{L(a_m)}{d}
  \int\limits_0^\infty \!d \gamma' P(\gamma')\, e^{-2 / \gamma'},
\label{D eps I}
\end{equation}
where $\gamma' = 1 / {\rm Im}\, \tau\omega_c$ and $P(\gamma')$ is
the distribution function of $\gamma'$. The fist factor on the
right-hand side is written solely to provide the correct dimensionality.

It is possible to show that $P(\gamma')$ has the Gaussian tail,
\begin{equation}
 P(\gamma') \sim \exp\left(-\frac{A \gamma'^2}{\gamma^2}\right),\quad
                 \gamma' \gg \gamma,
\label{P}
\end{equation}
where $A \sim 1$ is some number. This result can be obtained from
the following simple physical arguments. More rigorous treatment is
relegated to Appendix~\ref{proof}.

Let us again look back at the chessboard model. As one can see from
Eq.~(\ref{v_d}), $v_d$ as a function of $t$ exhibits a brief pronounced
pulse near its maximum at $t = t_0$. The duration of the pulse is of the
order of $(\gamma\omega_c)^{-1}$. It is this time scale that determines
the imaginary part of the closest singular point. Let us now return to
the random potential case. One can speculate that singular points of
$v_d(t)$ are always associated with such kind of pulses. By this
argument, the singularity at the point $t_s = t_1 - i t_2$ with $0 < t_2
\ll (\gamma\omega_c)^{-1}$ requires an unusually short pulse of duration
$\Delta t \sim t_2$. To produce such a pulse $v_d(s)$ must have a large
and sharp maximum. Let us estimate, e.g., the height of this maximum.
The half-width $\Delta s$ of the maximum is of the order of $\Delta s
\sim \sqrt{-v_d / v''_d}$. On the other hand, we should have $\Delta s
\sim v_d t_2$. Thus, $v_d v''_d \sim -t_2^{-2}$, which shows that small
values of ${\rm Im}\, t_s$ require large values of $v_d$ and its second
derivative, $v_d \sim d / t_2$ and $v''_d \sim 1 / t_2 d$. Recall now
that the distribution functions of both $v_d$ and $v''_d$ have Gaussian
tails, so that the probability of finding an unusually large $v_d$ is of
the order of $\exp(-A_1 v_d^2 / \gamma^2\omega_c^2 d^2)$ and similarly
for $v''_d$ ($A_1 \sim 1$ is some number). Substituting $d / t_2 \sim
\gamma'\omega_c d$ for $v_d$, we arrive at Eq.~(\ref{P}). The
calculation of $A$ for some particular example of $C(r)$ can be found in
Appendix~\ref{proof}.

The estimation of the integral in Eq.~(\ref{D eps I}) by the saddle-point
method results in
\begin{equation}
  \Delta\epsilon_m^2 \sim W_0^2 \frac{L(a_m)}{d}
  \exp\left(-\frac{3 A^{1/3}}{\gamma^{2/3}}\right).
\label{D eps II}
\end{equation}
On the other hand, $L(a_m)$ obeys the scaling law
\begin{equation}
            L(a_m) \propto |\Delta\epsilon_m|^{-\nu d_h},
\label{L_a}
\end{equation}
where $\nu$ and $d_h$ are some exponents, which depend on the properties
of the correlator $\widetilde{C}_0(q)$ [Ref.\onlinecite{Isichenko}].
Their actual values are not very important at this point.
Equations~(\ref{D eps II}) and (\ref{L_a}) enable one to find
$\Delta\epsilon_m$, which can then be substituted into Eq.~(\ref{D_de}).
As a result, we find the diffusion coefficient,
\begin{equation}
  D \sim \omega_c d^2 \gamma^\alpha
    \exp\left(-\frac{J}{\gamma^{2/3}}\right),
\label{D random}
\end{equation}
where $\alpha$ is some number and
\[
                 J = \frac{3 A^{1/3}}{1 + \nu d_h}
\]
is another number. We remind the reader that we cannot calculate the
correct pre-exponential factor in formula~(\ref{D random}). The
particular choice of this factor made in Eq.~(\ref{D random}) provides a
matching of this equation with Drude-Lorentz formula~(\ref{Drude}) at
$\gamma = 1$ where the both formulas give $D \sim \omega_c d^2$ (up to
purely numerical factors). This can be seen from Eqs.~(\ref{Drude}),
(\ref{gamma2}), and (\ref{D random}) if one takes into account the
approximate expression~\cite{Exact tau} for the transport time $\tau$,
\[
            \tau \sim \frac{d}{v}\left(\frac{E}{W}\right)^2.
\]

In this subsection we implicitly assumed that the inequality
$R_c \gg d$ holds. In this case $\gamma \propto B^{-3/2}$
[Eq.~(\ref{gamma2})]. Substituting this into Eq.~(\ref{D random}),
we obtain
\[
      D \propto \omega_c d^2 \gamma^\alpha e^{-B / B_c},\quad B > B_c.
\]
declared previously in Sec.~\ref{intro}. Note that the dependence of $D$
on the magnetic field is given by a simple exponential not the squeezed
one as in the chessboard model [Eq.~(\ref{D_of_B})]. The reason for
this difference comes from the important role of rare places on the
trajectories with unusually sharp features of the averaged potential
$U_0$.

\section{Inter-Landau level transition amplitudes}
\label{amplitudes}

In the preceeding Section we showed that in strong magnetic fields, $B >
B_c$, the guiding center of the cyclotron orbit closely follows the
level lines $U_0 = {\rm const}$ of the averaged potential $U_0$.
Nonvanishing diffusion coefficient appears due to small deviations from
the level lines. The characteristic value $\Delta\epsilon_m$ of such a
deviation was calculated purely classically. Due to the energy
conservation, $\Delta\epsilon_m$ also represents the change in the
kinetic energy $I\omega_c$ of the particle [Eq.~(\ref{Hamiltonian II})].

The purpose of this Section is to calculate the change in kinetic energy
quantum-mechanically by taking into account the discreetness of the
spectrum, i.e., the existence of the Landau levels (LLs). Note that this
is not yet a consistent quantum-mechanical treatment of the problem. For
example, in this Section we ignore localization and/or
quantum tunneling. An attempt to touch on some of those complicated
issues will be postponed till the next Section.

In quantum-mechanical terms, the change in kinetic energy results from
inter-LL transitions. Indeed, the change in kinetic energy due to $N
\to N + k$ transition is equal to $k\hbar\omega_c$. Denote the
transition amplitude upon the completion of the loop $U_0 = {\rm const}$
by $A_{N,N + k}$, then $\langle\Delta\epsilon_m^2\rangle$ is given by
\begin{equation}
\langle \Delta\epsilon_m^2\rangle = (\hbar\omega_c)^2 \sum_k k^2
|A_{N, N + k}|^2.
\label{Delta epsilon quantum}
\end{equation}
It is obvious from this formula that the inter-LL transitions may be
significant only within a certain band of LLs. If $\Delta\epsilon_m$ is
larger than $\hbar\omega_c$, then the number of LLs in that band should be
of the order of $\Delta\epsilon_m / \hbar\omega_c$. Denote by $B_\ast$
the field where $\Delta\epsilon_m = \hbar\omega_c$. In fact, this
notation has already been used in Sec.~\ref{intro} [Eq.~(\ref{B_ast})].
If $B > B_\ast$, then $\Delta\epsilon_m < \hbar\omega_c$ and even the
transitions to the neighboring LLs must be suppressed. In this case the sum
over $k$ is dominated by the two terms, $k = \pm 1$; therefore,
\begin{equation}
|A_{N, N \pm 1}|^2 =
\frac{\langle \Delta\epsilon_m^2\rangle}{2 (\hbar\omega_c)^2}.
\label{nearest}
\end{equation}
In deriving Eqs.~(\ref{Delta epsilon quantum}) and (\ref{nearest}) we
implicitly assumed that the classical and the quantum calculations of
$\langle \Delta\epsilon_m^2 \rangle$ give the same result. This will be
demonstrated below.

Before we do so, let us mention one interesting fact. Using
Eq.~(\ref{D_de}) and the Einstein relation $\sigma_{xx} = e^2 \nu D$
where $\nu = m / \pi \hbar^2$ is the density of states
(de~Haas-van~Alphen oscillations neglected) one arrives at the following
formula
\[
\sigma_{xx} \sim \frac{e^2}{h} \frac{\Delta\epsilon_m}{\hbar\omega_c}.
\]
It can be interpreted in the following way: the transport is determined
by the aforementioned band of about $(\Delta\epsilon_m / \hbar\omega_c)$
LLs with energies near the Fermi energy. Each level contributes
$\frac{e^2}{h}$ to $\sigma_{xx}$ [cf. Refs.~\onlinecite{E2_H}].

The general formula for $A_{N, N + k}$ derived in
Appendix~\ref{Transitions} reads
\begin{equation}
A_{N, N + k} = \int\limits_{0}^{2\pi} \frac{d \theta}{2 \pi}
e^{-i k \theta}
\exp\left( -\sum_{n \neq 0} \frac{\Delta_n}{n \hbar\omega_c}
e^{-i n \theta}\right),
\label{A_NM}
\end{equation}
where $\Delta_n$'s are given by Eq.~(\ref{Delta_n}).
Substituting this expression into formula~(\ref{Delta epsilon quantum})
and taking advantage of the identity
\[
\int\limits_0^{2\pi} \frac{d \theta}{2\pi} \sum_{k = -\infty}^{\infty}
k^2 e^{i \theta k} f(\theta) = -f''(0),
\]
we recover the classical formula~(\ref{Delta epsilon}) for
$\Delta\epsilon_m^2$.

Finally, it is easy to see that Eq.~(\ref{nearest}), which we derived
without any calculations, is consistent with formula~(\ref{A_NM}).
Indeed, $|\Delta_1| \simeq \Delta\epsilon_m$. If $\Delta\epsilon_m \ll
\hbar\omega_c$, then the second exponential in Eq.~(\ref{A_NM}) can be
expanded in the Taylor series, which trivially leads to
Eq.~(\ref{nearest}).


\section{Calculation of the quantum localization length}
\label{quantum}

In Sec.~\ref{intro} we argued that the localization length is
exponentially large in weak magnetic fields and has to decay as the
magnetic field increases. This statement is an oversimplification in two
respects. Firstly, $\xi$ is, in fact, expected to diverge at certain
discreet values $B_N$ of the magnetic field
\begin{equation}
      \xi = \xi_0 \left|\frac{B_{N + 1} - B_N}{B - B_N}\right|^{\mu},
\label{divergent xi}
\end{equation}
where $\mu$ is a critical exponent.~\cite{Prange} Secondly, such
divergences neglected, $\xi$ starts decreasing only from
$B \sim \hbar c / e l_{\rm tr}$, at which 
the magnetic length $l = \sqrt{\hbar / m\omega_c}$ becomes
of of the order of the transport length $l_{\rm tr} = v\tau$. 

Let us discuss these issues in some detail. Scaling theory of
localization is one possible way to approach this difficult
problem.~\cite{Lee_Rama} In scaling theory one tries to understand the
localization by considering the behavior of the dimensionless
conductance $g \equiv \frac{h}{e^2} \sigma_{xx}$ as a function of system
size $L$. This behavior is described by the scaling function
\begin{equation}
              \beta(g) = \frac{\partial \ln g}{\partial \ln L}.
\label{beta definition}
\end{equation}
One starts with calculating the conductance $g_0 = g(l_0)$ at some short
length scale $L = l_0$ where it is large and then finds how $g$
is renormalized towards larger $L$. The localization length is the
length scale where $g(L)$ becomes of the order of unity. (If
$g_0$ is of the order of unity of smaller, then a different approach
has to be used, see below).

It has been conjectured~\cite{Brezin} that all physical system can be
grouped into certain universality classes with the same functional form
of the scaling function. If we neglect the spin-orbit coupling, then the
appropriate universality class for our system is determined by the
relation between $L$ and the magnetic length $l$. For $L \ll l$, the
system belongs to the orthogonal class, where the scaling function is
given by~\cite{Lee_Rama,Brezin}
\begin{equation}
               \beta(g) \simeq -\frac{2}{\pi g},\quad L \ll l.
\label{beta orthogonal}
\end{equation}
For $L \gg l$ the system is in the unitary class. The scaling function
is given by
\begin{equation}
         \beta(g) \simeq -\frac{1}{2 \pi^2 g^2},\quad L \gg l.
\label{beta unitary}
\end{equation}
The latter result was derived both by the conventional diagram
technique~\cite{Carra,Salomon} and by an effective field
theory.~\cite{Brezin} Solving the scaling equation~(\ref{beta
definition}) for $g(L)$, we find that $\xi$ experiences a growth from
the value of
\begin{equation}
 \xi \sim l_{\rm tr} \exp\left(\frac{\pi}{2} k_F l_{\rm tr}\right)
\label{xi at zero}
\end{equation}
at $B = 0$ to
\begin{equation}
    \xi \sim l_{\rm tr} \exp\left(\pi^2 k_F^2 l_{\rm tr}^2\right)
\label{xi at small B}
\end{equation}
at $B \sim \hbar c / e l_{\rm tr}^2$ where $l = l_{\rm tr}$. In stronger
fields, $B > \hbar c / e l_{\rm tr}^2$, the system belongs to the
unitary class at all relevant length scales and $\xi$ is given by the
formula~\cite{Huckestein,Wei}
\begin{equation}
                       \xi = l_0 \exp[\pi^2 g_0(B)^2]
\label{xi}
\end{equation}
following from Eq.~(\ref{beta unitary}). The dimensionless conductance
$g_0(B)$ decreases with $B$. For the case of a long-range random
potential this follows from the results of the preceeding Sections.
Therefore, the initial {\em growth} of $\xi$ at very weak magnetic
fields is followed by the exponential {\em decay} of $\xi$ as $B$
increases. This is the statement we put forward in Sec.~\ref{intro}.
 
Unfortunately, Eq.~(\ref{xi}) cannot be entirely correct because it does
not reproduce the critical divergences [Eq.~(\ref{divergent xi})].
Pruisken~\cite{Pruisken} argued that the critical behavior is a
non-perturbative effect. His field-theoretical treatment yields an
expression for the $\beta$-function, in principle, different from the
simple form~(\ref{beta unitary}). However, the deviations from
Eq.~(\ref{beta unitary}) become significant only when the renormalized
value of $g$ approaches unity. On this basis we speculate that
Eq.~(\ref{xi}) gives only the lower bound for the localization length
$\xi$. We further assume that this lower bound is close to the actual
value of $\xi$ away from criticality. In other words, Eq.~(\ref{xi})
gives, in fact, not $\xi$ itself but its non-critical prefactor $\xi_0$
entering Eq.~(\ref{divergent xi}).

Note that $\xi \simeq \xi_0$ at the midpoints between neighboring
divergences of $\xi$, i.e., at the QHE conductivity minima. This is exactly
the quantity discussed in Sec.~\ref{intro} where we postulated the {\it
ansatz\/}~(\ref{xi_0}). We will rewrite it here for the ease of
reading:
\begin{equation}
              \xi_0 = l_0 \exp(\pi^2 g_0^2), \quad g_0 \gg 1.
\label{xi_0 II}
\end{equation}
By virtue of this {\it ansatz\/}, the calculation of $\xi_0$ boils down
to the evaluation of the short length scale conductance $g_0$. 

Previous attempts~\cite{Pruisken,Carra,Salomon} to treat the
localization problem in the QHE have been focused on the case of a
short-range random potential, i.e., the potential whose correlation
length is much smaller than de Broglie wavelength $2 \pi / k_F$. In this
case $g_0$ has to be calculated quantum-mechanically, e.g., within a
Self-Consistent Born Approximation (SCBA).~\cite{Ando,Laikhtman94} Recall
that our theory applies to the case $k_F d \gtrsim (E / W)^{2/3} \gg 1$,
see Eq.~(\ref{range}). Therefore, there is a whole intermediate region
$1 \ll k_F d \ll (E / W)^{2/3}$ separating the domains of applicability
of our and the previous theories. The calculation of $\xi_0$ in that
region is a separate problem and will be discussed elsewhere.

In the case of long-range random potential, which we consider here,
$g_0(B)$ can be calculated with the help of Einstein relation,
%
\[
                      g_0(B) = h \nu(B) D(B),
\]
%
where $\nu(B)$ is the density of states at the Fermi level and $D(B)$ is
the classical diffusion coefficient. According to the results of the
previous Sections, $D(B)$ is given by Drude-Lorentz
formula~(\ref{Drude}) at $B < B_c$ and by formula~(\ref{D random}) at $B
> B_c$. Let us now discuss the behavior of $\nu(B)$. In principle,
$\nu(B)$ oscillates with $B$ around its zero field value $\nu(0) = m /
\pi \hbar^2$. However, for $B$ smaller or at least not to much larger
than $B_c$ such oscillations are exponentially small because the width
of LLs, which is of the order of $W_0$ [Ref.\onlinecite{Raikh}], is much
larger than the distance $\hbar\omega_c$ between them. Therefore, we can
use the zero field value $\nu(0)$.
 
Substituting all these results into Eq.~(\ref{xi_0 II}), we obtain
$\xi_0(B)$. The functional form of this dependence is given by
Eqs.~(\ref{xi_0 weak}) and (\ref{xi_0 strong}). Graphically, it is
illustrated by Fig.~\ref{Plot xi}. Observe that the overall decay of
$\xi_0$ as $B$ increases becomes extremely sharp at $B > B_c$. As a
consequence, already at the field $B = B_\ast$, which is only
logarithmically larger than $B_c$[see Eq.~(\ref{B_c})] $\xi_0$ ceases to
be exponentially large. At $B > B_\ast$ $g_0$ becomes less than one and
Eq.~(\ref{xi_0 II}) does not hold any more. In this region the
localization length is determined mainly by quantum tunneling rather
than by the destructive interference of classical diffusion paths. Thus,
the calculation of $\xi_0$ requires a different approach. It will be
discussed in a forthcoming paper together with the prefactor in
formula~(\ref{xi_0 strong}). At this point we can only say that $\xi_0$ is
expected to have a power-law dependence on $B$ and eventually match
the predictions of Raikh and Shahbazyan~\cite{Raikh_xi} at sufficiently
large $B$.

\section{Discussion and conclusions}
\label{conclusions}

In this paper we studied a two-dimensional motion of a charged particle
in a weak long-range random potential and a perpendicular magnetic
field. We showed that the phase-space averaged diffusion coefficient is
given by Drude-Lorentz formula only at magnetic fields $B$ smaller than
certain value $B_c$. At larger fields, the chaotic motion is suppressed
and the diffusion coefficient becomes exponentially small.

To make connection with the experiment our results can be applied to the
following model. We suppose that the random potential is created by
randomly positioned ionized donors with two-dimensional density $n_i$
set back from the two-dimensional electron gas by an undoped layer of
width $d$. We will assume that $n_i d^2 \gg 1$ and also that $d \gg
a_B$, where $a_B$ is the effective Bohr radius. In this case the random
potential can be considered a Gaussian random potential whose correlator
is given in Appendix~\ref{uuu}. As a particular example, we consider a
special case where the density of {\it randomly positioned\/} donors is
equal to the density $k_F^2 / (2 \pi)$ of the electrons. We call it the
standard potential. It is easy to see that for the standard potential $E
/ W \sim k_F d$ and the domain of applicability of our theory
[Eq.~(\ref{range})] is simply $k_F d \gg 1$. In modern high-mobility
GaAs devices this parameter can be as large as ten. It is easy to verify
that the magnetic field $B_c$ where the classical localization takes
place corresponds to LL index $N_c \sim (k_F d)^{5/3}$, which can be a
number between $10$ and say, $50$ for GaAs heterostructures. Another
important magnetic field $B_\ast$ [Eq.~(\ref{B_ast})] corresponds to LL
index $N_\ast$, which is only slightly smaller than $N_c$. As explained
in Sec.~\ref{intro}, $N_\ast$ is the number of the ``first'' QHE plateau
in the sense that observability of plateaus with larger $N$ require {\em
exponentially small\/} temperatures.

The point $N = N_\ast$ plays another important role. It is the largest
$N$ where it is possible to see the activated transport $\sigma_{xx}
\propto e^{-E_a / T}$, $E_a \simeq \hbar\omega_c / 2$ in the
minima of $\sigma_{xx}$. Indeed, it is known that in strong fields or
for small $N$'s the dissipative conductivity demonstrates the
Arrenius-type behavior at not too low temperatures. As the temperature
decreases, the activation becomes replaced by the variable-range
hopping, see Eq.~(\ref{T_one_half}). 

Equating the two exponentials, we find the temperature $T_h$ at which the
activation gives in to the hopping,
\begin{equation}
                 T_h \sim \frac{(\hbar\omega_c)^2}{T_0}.
\label{hopping T}
\end{equation}
This formula can also be written in another form,
\begin{equation}
\frac{T_h}{\hbar\omega_c} \sim 
\frac{\hbar\omega_c}{T_0} = {\rm const}\, \frac{\xi_0}{r_s R_c},
\label{T_hw}
\end{equation}
where $r_s = \sqrt{2}\, e^2 / \kappa \hbar v_F$ is the gas parameter,
which is of the order of unity in practice. Let us demonstrate that the
activated behavior should not be observable at $B < B_\ast$. Indeed, it
make sense to talk about the activated behavior only at temperatures
below the activation energy $E_a \simeq \hbar\omega_c / 2$. Therefore,
the activated transport can be observable only if the right-hand side of
Eq.~(\ref{T_hw}) is less than unity. Thus, in the vicinity of $B =
B_\ast$ or near $N_\ast$-th conductivity minimum where $\xi_0$ suddenly
becomes exponentially large with decrease in $B$, there is no place for
the activated transport left. Strictly speaking, this argument only
proves that the activated behavior is absent at $B \lesssim B_\ast$.
However, it can be shown, and it is a subject of a forthcoming paper,
that the ratio $\xi_0(B_c) / R_c $ is smaller than one in the standard
case. Therefore, the activation indeed disappears at $B \simeq B_\ast$
rather than at some much stronger field.

The behavior of $\xi_0$ in magnetic fields stronger than $B_\ast$ has
not been investigated in the present paper. It will be discussed
elsewhere. We expect that at such magnetic fields $\xi_0(B)$ is a
certain power law matching the results of Raikh and
Shahbazyan~\cite{Raikh_xi} at sufficiently large $B$. As explained in
Sec.~\ref{intro}, such a dependence can be studied experimentally.

Finally, in this paper we have neglected the influence of
electron-electron interaction on $\xi_0$. This complicated issue warrants
further study.


\acknowledgements

We are grateful to A.~P.~Dmitriev, I.~V.~Gornyi, V.~Yu.~Kachorovskii,
A.~I.~Larkin, and D.~L.~Shepelyansky for useful discussions. This work
is supported by NSF under Grant DMR-9616880 and by the Russian Fund for
Fundamental Research.

\appendix

\section{Diffusion coefficient in the chessboard model}
\label{f_c}

To calculate the numerical factor in Eq.~(\ref{D I}) for the
diffusion coefficient we proceed as follows. First, we will introduce
the random-phase model~\cite{Ahn} arguing as follows. The well-known
property of the standard map is a fast mixing in the phase variable
$\vartheta$. The correlations in phase decay according to $\langle e^{i
(\vartheta_n - \vartheta_0)} \rangle \sim |K|^{-n / 2}$
[Ref.~\onlinecite{Zaslavsky}] as a function of the iteration number $n$;
therefore, for $|K| \gg 1$ the phase memory is typically lost after a
single iteration of the map. The situation with the separatrix map is
similar, which allows a simplification of the problem. We will assume
that $\epsilon_n$ is still transformed according to Eq.~(\ref{eps map})
as long as $|\epsilon_{n + 1}| \leq \epsilon_{\rm web}$. If the new
value of $|\epsilon_{n + 1}|$ is larger than $\epsilon_{\rm web}$, then
$\epsilon_{n + 1} = \epsilon_n$. At the same time, $\vartheta_n$ will be
a purely random variable uniformly distributed in the interval $(0,
2\pi)$. As we will see below, for transport only the narrow boundary
layer $|\epsilon| \sim \Delta \epsilon$ is important (cf.
Refs.~\onlinecite{Rosenbluth,Shraiman}) where $|K| \gg 1$ and therefore,
such a random-phase model is adequate. Ahn and Kim~\cite{Ahn}
studied this model numerically and found an excellent agreement between
the diffusion coefficients found from the random-phase model and from
the original separatrix map. (Of course, the random-phase model lacks
certain features of the original separatrix mapping, e.g., a rich
hierarchical island structure).

\begin{figure}
\centerline{
\psfig{file=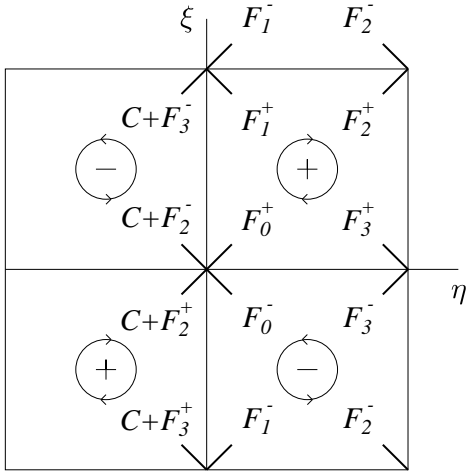,width=2.5in,bbllx=180pt,bblly=320pt,bburx=450pt,bbury=610pt}
}
\vspace{0.1in}
\setlength{\columnwidth}{3.2in}
\centerline{\caption{
The chessboard in the coordinate system rotated by $\frac{\pi}{4}$.
Pluses and minuses at the centers of the chessboard cells label maxima
and minima of the potential. The direction of the drift velocity is
indicated by arrows. Bold segments are the surfaces of section
$\Sigma_n^\epsilon$, the same as in Fig.~\protect\ref{sections}.
Distribution functions $F^\pm_n$ represent the deviation of the guiding
center density from the sample averaged value at those parts of
$\Sigma_n^\epsilon$'s, which are inside of the two cells on the right.
Surfaces $\Sigma_0^\epsilon$ and $\Sigma_1^\epsilon$ also penetrate the
two cells on the left. The corresponding distribution functions are
related to $F_2^\pm$, $F_3^\pm$ as shown. \label{diffusion}
}}
\end{figure}

Consider now an ensemble of particles, each having the same total energy
$E$ but different initial conditions at $t = 0$. In the original problem
with Hamiltonian~(\ref{Hamiltonian III}), we can describe this ensemble
by a distribution function (guiding center density) $F(\bbox{\rho}, t)$.
We will calculate the diffusion coefficient as the coefficient of
proportionality between the the average particle flux and the average
gradient of $F$ in the stationary state. It is convenient to rotate the
coordinate system by $\frac{\pi}{4}$. We denote new coordinates by $\xi$
and $\eta$. The gradient of $F$ is in the $\hat{\bbox{\eta}}$-direction
(Fig.~\ref{diffusion}).

%
%

In fact, the description of the ensemble by function $F$, which is a
function of a vector argument, is reasonable only when we study the
exact dynamics. After we have replaced the exact dynamics with that of
the separatrix map and now even of the random-phase model, this kind of
description became too detailed. Instead, it is sufficient to introduce
a set of the distribution functions $F^\pm_n(\epsilon)$ of a single
argument. Each function in the set represents the deviation of $F$ from
its average value at the intersections of the contour $U_0 = \epsilon$
with the surfaces of section $\Sigma^\epsilon_n$. The superscripts
distinguish between the positive and negative $\epsilon$ contours
(Fig.~\ref{diffusion}). Functions $F^+_n$ are taken to be zero for
$\epsilon < 0$ and similarly, $F^-_n(\epsilon) = 0$ for $\epsilon > 0$.
We also define ``full'' functions $F_n$ by $F_n(\epsilon) =
F^+_n(\epsilon) + F^-_n(\epsilon)$.

Denote the length of the chessboard cells by
$d_0$ ($d_0 = \sqrt{2}\,\pi d$) and the average gradient $\langle|\nabla
F|\rangle$ by $\Delta F / d_0$, then
\begin{equation}
            D = \frac{\Phi}{\Delta F} = \frac{\Phi_1 - \Phi_0}{\Delta F},
\label{D from f}
\end{equation}
where $\Phi$ is the total flux through the side $\{0 < \xi < d_0,\:\:
\eta = 0\}$ and $\Phi_n$ is the flux incident upon the $\Sigma^\epsilon_n$
surface,
\begin{equation}
\Phi_n = \int\limits_0^{\pi d} d l\, v_d(l)\, F^+_n[\epsilon(l)]
      = \frac{1}{m\omega_c}\int\limits_0^{W_0} d \epsilon F^+_n(\epsilon).
\end{equation}
Here $l$ is the coordinate along $\Sigma^\epsilon_n$ and $v_d(l)$ is the
drift velocity.

\begin{figure}
\centerline{
\psfig{file=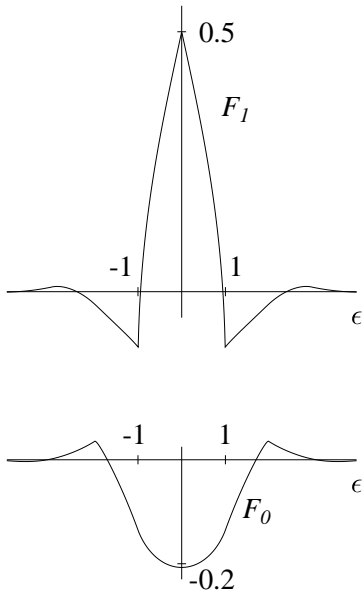,width=2.1in,bbllx=200pt,bblly=280pt,bburx=435pt,bbury=635pt}
}
\vspace{0.1in}
\setlength{\columnwidth}{3.2in}
\centerline{\caption{
The distribution functions $F_0$ and $F_1$. The density (vertical axis)
is in units of $\Delta F$ and the energy (horizontal axis) in units of $\Delta
\epsilon$.
\label{F_1}
}}
\end{figure}

To obtain the equation for $F_n$'s note that within the random-phase
model, $F_{n + 1}$ is quite simply related to $F_n$. For example,
\begin{equation}
 F_2(\epsilon) = \int\limits_0^{2\pi}\frac{d \vartheta}{2\pi}
 F_1(\epsilon - \Delta \epsilon\sin\vartheta) = S \circ F_1(\epsilon),
\label{S1}
\end{equation}
Similarly (see Fig.~\ref{diffusion}),
\begin{eqnarray}
 \Delta F \Theta(-\epsilon) &+& F^-_3(\epsilon) + F^+_1(\epsilon)\nonumber\\ 
&=& S \circ [\Delta F \Theta(-\epsilon) + F^-_2(\epsilon) + F^+_0(\epsilon)],
\label{S2}
\end{eqnarray}
where $\Theta(x)$ is the step-function. Suppose that all $F_n$'s are
equal to zero at the center of the cell, then the chessboard symmetry
dictates $F_3 = - F_1$ and $F_2 = - F_0$ and also that functions
$F_n$'s are even. These relations can be substituted into Eq.~(\ref{S2}).
Then one can eliminate $F_0$ and obtain an equation solely for $F_1$,
\begin{equation}
(1 + I \circ S \circ I \circ S) F_1(\epsilon) =
(I \circ S - I) \Delta F \Theta(-\epsilon),
\label{Eq for F_1}
\end{equation}
where $(I \circ f)(\epsilon) \equiv {\rm sgn}(\epsilon) f(\epsilon)$.
Equation~(\ref{Eq for F_1}) is an integral equation, in principle
solvable by the Winer-Hopf method. However, we have not been able to
find its solution analytically. At the same time, a numerical solution
can be obtained rather easily. The result is shown in Fig.~\ref{F_1}. By
calculating the area bounded from above by the graph of function $F_1$
from below by the graph of $F_0$ and from the left by the vertical line
$\epsilon = 0$ we have obtained the numerical factor $0.45$ in the
expression~(\ref{D I}) for the diffusion coefficient.

Both $F_0(\epsilon)$ and $F_1(\epsilon)$ decay exponentially at
$|\epsilon| \gg \Delta \epsilon$. This is in accordance with the
statement above that only a narrow boundary layer is important for the
transport. Similar to the conventional advection-diffusion
problems,~\cite{Rosenbluth,Shraiman} the width of this layer, $d \Delta
\epsilon / W_0$, is of the order of the average displacement of the particle
in the direction perpendicular to the flow upon travelling the length of
the chessboard's cell.

%
%
%
\section{Asymptotic behavior of function $P({\gamma'})$}
\label{proof}

In Sec.~\ref{strong} we showed that the calculation of the diffusion
coefficient in strong magnetic fields can be reduced to the calculation
of the distribution function $P(\gamma')$. Our derivation of the
subsequent formula~(\ref{D random}) was based on Eq.~(\ref{P}), which we
rewrite here for convenience,
\begin{equation}
    P(\gamma') \propto \exp\left(-\frac{A \gamma'^2}{\gamma^2}\right).
\label{P I}
\end{equation}
Our goal is to derive Eq.~(\ref{P I}).

The sufficient conditions for Eq.~(\ref{P I}) to hold seem to be as
follows. Suppose that $C(\bbox{r})$ is isotropic, i.e., depends only on
$r = \sqrt{x^2 + y^2}$. Function $C(r)$ must be analytic for all real
$r$. In addition, $C(r)$ must be analytic in some complex
neighborhood of $r = 0$. Note that such conditions can be met only if
$\widetilde{C}(q)$ decays sufficiently fast at large $q$. The
``realistic'' potential [Eqs.~(\ref{C_q real}) and (\ref{C_r real})]
meets the aforementioned requirements. It is easy to see, for instance,
that in this particular example $C(r)$ is analytic within the circle
$|r| < 2 d$ in the complex plane of $r$.

Below we use the symbol $v(s)$, which is a concise notation for the
drift velocity on the appropriate closed-loop trajectory. The argument
$s$ has the meaning of the coordinate along this trajectory. Let us
introduce the following definition. Consider an arbitrary point $r$ on
the trajectory. We call the function $u(s)$,
\[
u(s)\equiv\sum_{n = 0}^\infty\frac{\left|v^{(n)}(r)\right|}{n!}(s - r)^n
\]
the majorant of $v(s)$ at the point $r$. The derivation of Eq.~(\ref{P I})
is based on the following\hfill\\[-4pt]

\noindent{\bf Lemma} {\it Suppose that function $v(s)$ is analytic in a
complex neighborhood of point $r$. Consider the solution $s(t)$ of
the Cauchy problem}
\begin{equation}
          \frac{d s}{d t} = v(s),\quad s(t_0) = r
\label{Cauchy}
\end{equation}
{\it as a function of the complex argument $t$ and define $v(t) \equiv
v[s(t)]$. In the similar way we define function $u(t)$ where $u(s)$
is the majorant of $v(s)$ at point $r$.

Under such conditions the shortest distance $t_v$ from $t_0$ to
a singular point of $v(t)$ is greater or equal to the similarly
defined distance $t_u$ for the function $u(t)$.}

\newtheorem{corollary}{Corollary}
\begin{corollary}
$t_v$ satisfies the inequality
\begin{equation}
t_v \geq \int\limits_r^{r_m(r)}
\frac{d s}{a_0 + a_1 (s - r) + a_2 (s - r)^2 +\ldots},
\label{t_v estimate}
\end{equation}
where
\[
             a_n = \frac{\left|v^{(n)}(r)\right|}{n!},
\]
and $r_m(r)$ is the smallest real number larger than $r$
such that $u(r_m) = \infty$.
\label{cor1}
\end{corollary}

{\it Proof.\/} The distance $t_v$ can be calculated as
\begin{equation}
t_v = \liminf\limits_{n \to\infty}
   \left|\frac{n! \,v_0}{\left(\frac{d}{d t}\right)^n \! v(t)}\right|^{1/n},
\label{t_v}
\end{equation}
where the derivatives are taken at $t = t_0$ and $v_0$ is an arbitrary
positive parameter with dimension of velocity. A similar expression can
be written for $t_u$. Clearly,
\[
\frac{d^n}{d t^n} \left.v(t)\right|_{t = t_0}
= \left(v \frac{d}{d s}\right)^n \left.v(s)\right|_{s = r}.
\]
Upon taking the derivatives, the right-hand side becomes
a sum of products of $v^{(k)}(s)$. It is important that
all the coefficients in the sum are positive, which leads to the
inequality
\[
\left|\frac{d^n v}{d t^n}\right| \leq
\left|\frac{d^n u}{d t^n}\right|.
\]
The statement of the Lemma follows from this inequality
and Eq.~(\ref{t_v}).

To prove Corollary~\ref{cor1} note that the differential equation on
$s(t)$ with $u(s)$ on the right-hand side can be solved in quadratures,
\[
            t(s) - t_0 = \int\limits_r^{s} \frac{d s}{u(s)}.
\]
At singular points $u(t)$ diverges; therefore, the distance to such a
point can be found by choosing the upper limit from the condition
$|u(s)| = \infty$. It is easy to see that by choosing $r_m$ for the
upper limit and the integration path along the real axis, we find the
shortest distance $t_u$ to the singularity. Inequality~(\ref{t_v
estimate}) trivially follows from this result~\rule{0.07in}{0.045in}

Recall that we are interested in finding the imaginary parts of
the singular points. For this purpose the following Corollary
is helpful.
\begin{corollary}
Suppose that $v(s)$ is a periodic function with period $L$. The
imaginary part ${\rm Im}\, t_s$ of any singular point $t_s$ of function
$v(t)$ defined by~{\rm(\ref{Cauchy})} satisfies the inequality
\begin{equation}
{\rm Im}\, t_s \geq \min\limits_{0 < r < L} \int\limits_0^{r_m(r) - r}
\frac{d s}{a_0 + a_1 s + a_2 s^2 +\ldots}.
\label{Im t_v estimate}
\end{equation}
\label{cor2}
\end{corollary}
The proof is quite obvious.

At this point we are ready to derive Eq.~(\ref{P I}). The derivation is
based on the rigorous bound~(\ref{Im t_v estimate}) supplemented by a
few quite plausible assumptions. Consider some contour $U_0(s) = {\rm
const}$ with perimeter length $L$ and choose a point $s = r$ on this
contour. All $a_n$'s entering~(\ref{Im t_v estimate}) are random numbers
with certain distributions such that $a_n$ larger than some
characteristic values $A_n$ are exponentially rare. Generally, $A_n$ is
of the order of the typical value of $(n + 1)$-th order derivative of
$U_0(\bbox{\rho}) / m\omega_c$ with respect to $\rho_x$ and/or $\rho_y$,
e.g., $\partial^{n + 1} U_0(\bbox{\rho}) / \partial \rho_x^{n + 1}$. The
latter quantity has the normal distribution with variance $O(n^{-1})
\left|C_0^{(2 n + 2)}(0)\right|$. Hence,
\[
               A_n \sim n^\alpha \left|C_0^{(2 n + 2)}(0)\right|,
\]
where $\alpha$ is some number.

Let $\rho_m$ be the shortest distance from the origin to the singular
point of $C_0(\rho)$ in the complex plane of $\rho$. We will assume that
$\rho_m < \infty$ as in our example [Eq.~(\ref{C_r real})] where
$\rho_m = 2 d$. (The
analysis of the other case, $\rho_m = \infty$ is similar). Based on this
example, we further assume that $\rho_m \sim d$.

Since $\rho_m$ is the radius of the analyticity circle of $C_0(\rho)$,
we must have [cf. Eq.~(\ref{t_v})]
\[
\limsup\limits_{n \to\infty}
\left|\frac{1}{n!}\frac{C_0^{(n)}(0)}{C_0(0)}\right|^{1 / n} =
\frac{1}{\rho_m}.
\]
Consequently, the asymptotic behavior of $A_n$ at large $n$ is given by
\begin{equation}
\left(\frac{A_n}{\gamma\omega_c d}\right)^{1/n} \sim \frac{1}{\rho_m}
\sim \frac{1}{d}.
\label{A_n}
\end{equation}
The typical value of $a_n$ in formula~(\ref{Im t_v estimate}) is of the
order of $A_n$; hence, the typical distance of singular points of $v(t)$
from the real axis is of the order of $(\gamma \omega_c)^{-1}$. This is
the statement we put forward in Sec.~\ref{random}. Next we would like to
identify the condition for $v(t)$ to have a singular point $t_s$ with
$|{\rm Im}\, t_s| \ll (\gamma\omega_c)^{-1}$. It is more convenient to
work with dimensionless variable $\gamma' \equiv ({\rm Im}\, t_s
\omega_c)^{-1}$ introduced in Sec.~\ref{random}. The typical value of
$\gamma'$ is of the order of $\gamma$. We want to estimate the
probability density $P(\gamma')$ of large $\gamma'$, $\gamma' \gg
\gamma$. According to Corollary~\ref{cor2}, large $\gamma'$ may occur
only if $a_k \gg A_k$ for some $k$'s. Suppose for a moment that the
number $n$, which is the smallest of such $k$'s, is larger than two. Let
us estimate how large the corresponding $a_n$ should be to produce a
given value of $\gamma'$. It is easy to see that $(a_n / a_0)^{1 / n}
\gtrsim \gamma'\omega_c / a_0$ is needed. In view of Eq.~(\ref{A_n}),
this can be written as $(a_n / A_n)^{1 / n} \sim \gamma' / \gamma$. On
the other hand, the probability of having a large $a_n$ is, roughly
speaking, $\exp(-a_n^2 / A_n^2)$. This reasoning shows that it is
extremely inefficient to produce large $\gamma'$ by boosting $a_n$ with
$n > 2$. Therefore, the asymptotic behavior of function $P(\gamma')$
can be obtained by assuming that all $a_n$ in~(\ref{Im t_v estimate})
with $n > 2$ have typical values and therefore, can be omitted, while
$a_0$, $a_1$, and $a_2$ may be large. (This is the smallest number of
terms needed for the convergence of the remaining integral). It is also
quite straightforward to verify that large $a_1$'s are not as efficient
for producing small ${\rm Im}\, t_s$'s as the large values of $a_0$ and
$a_2$. Hence, $a_1$ may be omitted as well. Thus, we arrive at the
estimate
\[
\frac{1}{\gamma'\omega_c} \geq \int\limits_0^{\infty}
\frac{d s}{a_0 + a_2 s^2} = \frac{\pi}{2 \sqrt{a_0 a_2}},
\]
which is the same as
\[
\frac{1}{\gamma'\omega_c} \geq \frac{\pi}{\sqrt{2 |v v''|}}.
\]
To obtain the asymptotic behavior of the distribution function
$P(\gamma')$ we can replace the inequality sign by equality, and so
\[
P(\gamma') \simeq \int\limits_0^\infty d v
                \int\limits_{-\infty}^\infty d v''
\delta\left(\gamma' - \frac{\sqrt{2 |v v''|}}{\pi\omega_c}\right)
{\rm Prob}\,(v, v''),
\]
where ${\rm Prob}\,(v, v'')$ is the joint distribution function of $v$
and $v''$. Neglecting the terms important only for the pre-exponential
factors, we arrive at
%
\[
P(\gamma') \propto {\rm Prob}\,
\left[v_s, -\frac{(\pi\omega_c\gamma')^2}{2 v_s}\right],
\]
%
where $v_s$ is the saddle-point defined by the equation
\[
\frac{d}{d v_s}\ln {\rm Prob}\,
\left[v_s, -\frac{(\pi\omega_c\gamma')^2}{2 v_s}\right] = 0,
\]
Equation~(\ref{P I}) can be obtained from here taking
into account the Gaussian decay of the joint distribution function, $\ln {\rm
Prob}\,(v, v'') \sim -(v / A_0)^2 - (v'' / A_2)^2$ for $v \gg A_0$ and
$|v''| \gg A_2$. In this simplified model $v_s \sim \gamma'\omega_c d$
and Eq.~(\ref{P I}) follows.

Finally, let us calculate the numerical factor $A$, which enters
Eq.~(\ref{P I}), for the case of $C(r)$ given by Eq.~(\ref{C_r real}).
First we express $v$ and $v''$ in terms of the derivatives of $U_0$. For
the sake of notation simplicity we will drop the subscript ``0'' of
$U_0$. In addition, we denote the partial derivatives with respect to
$\rho_x$ by subscript $x$ and with respect to $\rho_y$ by $y$. It is
easy to see that
%
\[
        v = \frac{\sqrt{U_{x}^2 + U_{y}^2}}{m\omega_c}
\]
%
and
%
\[
v'' = \left(
\frac{U_{x}\frac{\partial}{\partial\rho_y} -
      U_{y}\frac{\partial}{\partial\rho_x}}{\sqrt{U_{x}^2 + U_{y}^2}}
\right)^2 v(\bbox{\rho}).
\]
%
Consider the list $\frac{1}{m\omega_c}\{U, U_x, U_y, U_{xx},\ldots,
U_{yyy} \}$ of $U$ and its derivatives arranged in lexicographical
order. We will refer to the members of this list as $u_0, u_1,\ldots,
u_9$, respectively. After some algebra, one can derive the following
results,
\[
                      v = \sqrt{u_1^2 + u_2^2}
\]
and
\begin{eqnarray*}
 v'' v^5 &=&
  -u_2^2\,u_5^2\, u_1^2 + 
   u_2^3\, u_9\, u_1^2 + 
   u_5^2\, u_1^4 + 
   u_2\, u_9\, u_1^4 \\&+&
   2\, u_2^3\, u_5\, u_1\, u_4 - 6\,u_2\, u_5\,
   u_1^3\, u_4 + 
   8\, u_2^2\, u_1^2\, u_4^2 \\&-& 
   2\, u_2^4\, u_1\, u_8 -
   u_2^2\, u_1^3\, u_8 + 
   u_1^5\, u_8 - 
   u_2^4\, u_5\, u_3 \\&+& 
   2\, u_2^2\, u_5\, u_1^2\, u_3 -
   u_5\, u_1^4\, u_3 - 
   6\, u_2^3\, u_1\, u_4\, u_3 \\&+&
   2\, u_2\, u_1^3\, u_4\, u_3 + 
   u_2^4\, u_3^2 -
   u_2^2\, u_1^2\, u_3^2 + 
   u_2^5\, u_7 \\&-& 
   u_2^3\, u_1^2\, u_7 - 
   2\, u_2\, u_1^4\, u_7 +
   u_2^4\, u_1\, u_6 \\&+& 
   u_2^2\, u_1^3\, u_6.
\end{eqnarray*}
Using the fact that $u_k$'s are Gaussian random variables, we obtain
\begin{eqnarray}
\displaystyle P(\gamma') &\propto& \int\prod_{k = 0}^{9} d u_k\exp
\left[-\frac12(\bbox{u}, { K^{-1}} \bbox{u})\right]
\nonumber\\
\displaystyle \mbox{} &\times& 
\delta\left(\gamma' - \frac{\sqrt{2 |v v''|}}{\pi\omega_c}\right),
\label{P Gauss}
\end{eqnarray}
where $K$ is $10 \times 10$ matrix with elements
\[
             K_{m n} = \langle u_m u_n \rangle.
\]
This matrix can be expressed in terms of derivatives of
$C_0(\bbox{\rho})$ at the point $\bbox{\rho} = 0$.

The application of the saddle-point method to the integral in
Eq.~(\ref{P Gauss}) leads to the estimate
\[
       P(\gamma') \propto e^{-\min P_1(\bbox{u})},
\]
where
\[
P_1(\bbox{u}) = -\frac12 \left(\bbox{u}, { K^{-1}} \bbox{u}\right),
\]
and the minimum is sought under the condition
\[
P_2(\bbox{u}) \equiv \sqrt{|v v''|} = \frac{\pi\gamma'\omega_c}{\sqrt{2}}.
\]
In fact, we exercise another refinement requiring that $u_0 = 0$,
which reflects the fact that we are interested in contours near the
percolation level.

The formulated minimization problem can by solved by Lagrange's
multiplier method, which requires the minimization of the form $P_1 -
\lambda P_2$. It is easy to see, however, that it can in its turn be
reduced to the (unrestricted) minimization of the function $P_1(\bbox{u}) -
P_2(\bbox{u})$. Let $\bbox{u}_*$ be the solution of the latter problem,
then the number $A$ appearing in Eq.~(\ref{P I}) is given by
\[
A = \frac{(\pi\gamma\omega_c)^2}{2}
    \frac{P_1(\bbox{u}_*)}{P_2^2(\bbox{u}_*)}.
\]
Our result for $C$ from Eq.~(\ref{C_q real}) is
\begin{equation}
                          A = 4.998.
\label{A}
\end{equation}
%

\section{Realistic random potential}
\label{uuu}

It has been suggested that a good model for the random potential really
existing in ${\rm GaAs}$ devices is the following one:
\begin{equation}
          \widetilde{C}(q) = 8 \pi W^2 d^2 e^{-2 q d},
\label{C_q real}
\end{equation}
or equivalently,
\begin{equation}
           C(r) = \frac{W^2}{(1 + r^2 / 4 d^2)^{3/2}}.
\label{C_r real}
\end{equation}

Equations~(\ref{C_q real}) and (\ref{C_r real}) corresponds to the potential
created in the plane of the two-dimensional electron gas by randomly
positioned ionized donors set back by an undoped layer of width $d$. The
amplitude of the potential has the following relation to the parameters
of the heterostructure
\begin{equation}
             W^2 = \frac{\pi}{8} \frac{n_i (e^2 a_B)^2}{d^2},
\label{W_real}
\end{equation}
where $n_i$ is the density of the donors and $a_B$ is the effective Bohr
radius of the electron gas. Equation~(\ref{W_real}) applies provided $d
\gg a_B$. The random potential can be considered a Gaussian random
potential if $n_i d^2 \gg 1$.

Using Eq.~(\ref{C_q real}) for the bare potential, one
can also obtain the real-space correlator $C_0(\rho)$ of the averaged
potential. Let $R_c \gg d$, then the following relations hold,
\begin{eqnarray*}
\displaystyle C_0(\rho) &\simeq& W_0^2 \left(1 - \frac{\rho^2}{8 d^2}\right),
\:\: 0 \leq \rho \lesssim d,\\
\displaystyle\mbox{} &\simeq& W_0^2
\frac{4 R_c d}{\rho\sqrt{4 R_c^2 - \rho^2}},\:\:
\rho \gtrsim d \:\,\,{\rm and}\:\,\, 2 R_c - \rho \gtrsim d,\\
\displaystyle\mbox{} &\simeq& W^2 \frac{8 d^3}{\rho^3}
\left(1 + \frac92\frac{R_c^2}{\rho^2}\right),\:\: \rho \gg 2 R_c,
\end{eqnarray*}
where
\[
                    W_0 = W \sqrt{\frac{2 d}{\pi R_c}}.
\]
Note that the $1 / \rho$ decay of $C_0(\rho)$ for $d \ll \rho \ll R_c$
is a universal feature of $C_0(\rho)$.

\section{Calculation of quasiclassical transition amplitudes}
\label{Transitions}

To derive Eq.~(\ref{A_NM}) we start with quantizing the classical
Hamiltonian~(\ref{Hamiltonian II}). The result is
\[
\hat{H} = \frac{m \omega_c^2}{2} (\hat{R}_x^2 + {R}_y^2)
  + U(\hat{\rho}_x + \hat{R}_x, {\rho}_y + {R}_y),
\]
where hats are used to denote the operators, {\it viz.\/},
\[
\hat{R}_x = i l^2 \frac{\partial}{\partial R_y},\:\:
\hat{\rho}_x = -i l^2 \frac{\partial}{\partial \rho_y},
\]
with $l = \sqrt{\hbar / m\omega_c}$ being the magnetic length. However,
since the guiding center motion is slow and quasiclassical, we can treat
$\rho_x$ as a classical dynamic variable with the equation of
motion~(\ref{rho_x_dot}) and similarly for $\rho_y$.

As in Sec.~\ref{strong}, we replace the exact trajectory
$\bbox{\rho}(t)$ by the ``unperturbed'' one, $\bbox{\rho}_0(t)$.
Everything, which was said in Sec.~\ref{strong} about the validity of
such an approximation, applies here as well.

The Schr\"oedinger equation
\begin{equation}
 i\hbar \frac{\partial}{\partial t} \Psi(R_y, t) = \hat{H} \Psi(R_y, t),
\label{Schroedinger}
\end{equation}
where from now on
%
\[
\hat{H} = \frac{m \omega_c^2}{2} (\hat{R}_x^2 + {R}_y^2)
  + U[\rho_{0x}(t) + \hat{R}_x, \rho_{0y}(t) + {R}_y]
\]
%
describes the evolution of the cyclotron motion under the influence of
time-dependent perturbation $\hat{U}(t)$. The solution of
Eq.~(\ref{Schroedinger}) will be sought in the form
\begin{equation}
\Psi(R_y, t) = \sum_{M = 0}^\infty c_M(t)
\Phi_M^0(R_y)\, e^{-i \left(M + \frac12\right) \omega_c t},
\label{Psi}
\end{equation}
where function $\Phi_M^0(R_y)$, given by
\[
\Phi_M^0(R_y) = \frac{1}{\sqrt{2^M M!\,l}\,(\pi)^{1/4}}\,
e^{-\frac{R_y^2}{2 l^2}} {\cal H}_M \left(\frac{R_y}{l}\right),
\]
represents the unperturbed wavefunction of $M$-th LL (${\cal H}_M$ is
the Hermite polynomial). Using this expression, one can find the matrix
elements $U_{M, M + k}(t) = \langle M | \hat{U}(t) | M + k\rangle$.
If $M$ is large and $|k| \ll M$, then it is sufficient to use the
quasiclassical approximation [cf. Ref.\onlinecite{Landau}, Sec.~51]
\[
        U_{M, M + k}(t) \simeq U_k[\bbox{\rho}_0(t)],
\]
where $U_k$ is the Fourier coefficient defined by Eq.~(\ref{U_k}). With
the help of this approximation, the equation for the expansion
coefficient $c_M$ can be written as follows
%
\[
i\hbar \frac{d c_M}{d t} = \sum\limits_{k = -\infty}^{\infty}
c_{M + k}\, U_k[\bbox{\rho}_0(t)]\, e^{-i k \omega_c t}.
\]
%
It has the solution
\[
c_M = \int\limits_{0}^{2\pi}\, \frac{d \theta}{2 \pi}
\lambda_0(\theta)\,
e^{-i M \theta}
\exp\left( \sum_k S_k(t) e^{-i k \theta}\right),
\]
where $\lambda_0(\theta)$ depends on the initial conditions at $t = t_0$
and
\[
S_k(t) = -\frac{i}{\hbar} \int\limits_{t_0}^t \!
U_k(t') e^{-i k \omega_c t'} d t'.
\]

To elucidate the structure of this solution note that
formula~(\ref{Psi}) can be rewritten in the form
\[
\Psi(R_y, t) = \sum_{n} b_n(t)\, \Phi_n(R_y, t)\,
e^{-\frac{i}{\hbar} E_n t},
\]
where the relation of $\Phi_n(R_y, t)$'s to $\Phi_M^0(R_y)$'s is as follows
\[
\Phi_n = \sum_M \Phi_M^0 \!
\int\limits_{0}^{2\pi} \! \frac{d \theta}{2 \pi}\,
e^{i (n - M) \theta} \exp \! \left[\sum_{k \neq 0}
\frac{U_k(t)}{k \hbar \omega_c} e^{-i k (\omega_c t + \theta)}\right].
\]
The new expansion coefficients $b_n$'s are given by
\begin{equation}
b_n = \int\limits_{0}^{2\pi}\! \frac{d \theta}{2 \pi}
\lambda(\theta)\,
e^{-i n \theta}
\exp\!\left[ -\sum_{k \neq 0} \frac{\Delta_k(t_0, t)}{k \hbar\omega_c}
e^{-i k \theta}\right],
\label{b_n}
\end{equation}
where initial conditions now enter through function $\lambda(\theta)$
and $\Delta_k(t_0, t)$ denotes the following integral
\[
\Delta_k(t_0, t) = \int\limits_{t_0}^t \! \dot{U}_k[\bbox{\rho_0}(t')]
e^{-i k \omega_c t'} d t'.
\]
Functions $\Phi_n(R_y, t)$ represent the ``instantaneous'' LL functions at a
given point $\bbox{\rho}_0(t)$ on the drift trajectory. They are the
eigenfunctions of $\hat{H}$ with a ``frozen'' value of $\bbox{\rho}_0$.
The corresponding eigenvalues $E_n$, however, turn out to be
time-independent $E_n = \left(n + \frac12\right)\hbar\omega_c + U_0$.
The direct physical meaning have the transitions between the
instantaneous states $\Phi_n$ not between the unperturbed states
$\Phi_M^0$. It is the former transition amplitudes we are going to
calculate (see a similar discussion in Ref.\onlinecite{Landau}, Sec.
41).

After Eq.~(\ref{b_n}) is obtained, we can choose any initial conditions,
for instance, $\lambda(\theta) = e^{i N \theta}$ such that $b_n(t_0) =
\delta_{n, N}$. In this case $b_{N + k}(t)$ gives the desired $N \to N +
k$ inter-LL transition amplitude, i.e., $A_{N, N + k}$
[Eq.~(\ref{A_NM})].


\end{multicols}

\begin{references}
\bibitem{Comment on Treschev} Our Hamiltonian
contains an infinite number of $U_k$'s, which are all different.
This makes our problem much more complicated than the two treated by
Treschev:

{\sl Problem 1}. $U_k = 0,\quad |k| > 1$, and

{\sl Problem 2}. $U_k = U_{-k}^* = U_1, \quad k > 0$.
\end{references}
\end{document}